\documentclass[fleqn,usenatbib]{mnras}

\usepackage{newtxtext,newtxmath}

\usepackage[T1]{fontenc}

\DeclareRobustCommand{\VAN}[3]{#2}
\let\VANthebibliography\thebibliography
\def\thebibliography{\DeclareRobustCommand{\VAN}[3]{##3}\VANthebibliography}

\usepackage{graphicx}	
\usepackage{amsmath}	
\usepackage{amssymb}	
\usepackage{color}

\graphicspath{{./}{figures/}}

\title[NGC 247 young faint fuzzy star cluster]{Low-density star cluster formation: discovery of a young faint fuzzy on the outskirts of the low-mass spiral galaxy NGC 247}

\author[A. J. Romanowsky et al.]{Aaron J.\ Romanowsky,$^{1,2,3}$\thanks{E-mail: aaron.romanowsky@sjsu.edu (AJR)}
S{\o}ren S.\ Larsen,$^{4}$
Alexa Villaume$^{5,1,3}$, 
Jeffrey L.\ Carlin$^{6}$,
Joachim Janz$^{7,8,9}$,
\newauthor
David J.\ Sand$^{10}$, 
Jay Strader$^{11}$, 
Jean P.\ Brodie$^{12,2}$,
Sukanya Chakrabarti$^{13,14}$,
Chloe M.\ Cheng$^{5}$, 
\newauthor
Denija Crnojevi\'c$^{15}$, 
Duncan A.\ Forbes$^{12}$, 
Christopher T.\ Garling$^{16}$, 
Jonathan R.\ Hargis$^{17}$, 
\newauthor
Ananthan Karunakaran$^{18}$, 
Ignacio Mart\'in-Navarro$^{19,20,2}$, 
Knut A.G.\ Olsen$^{21}$,
Nicole Rider$^{22}$, 
\newauthor
Bitha Salimkumar$^{1}$, 
Vakini Santhanakrishnan$^{1}$, 
Kristine Spekkens$^{23}$,
Yimeng Tang$^{3}$, 
\newauthor
Pieter G.\ van Dokkum$^{24}$ 
and Beth Willman$^{21}$ 
\\
$^{1}$Department of Physics \& Astronomy, San Jos\'e State University, One Washington Square, San Jose, CA 95192, USA\\
$^{2}$University of California Observatories, 1156 High Street,
Santa Cruz, CA 95064, USA\\
$^{3}$Department of Astronomy \& Astrophysics, University of California Santa Cruz, 1156 High Street, Santa Cruz, CA 95064, USA\\ 
$^{4}$Department of Astrophysics/IMAPP, Radboud University, 
PO Box 9010, 6500 GL Nijmegen, The Netherlands\\
$^{5}$Waterloo Centre for Astrophysics, Department of Physics \& Astronomy, University of Waterloo, 200 University Ave.\ W., Waterloo, Ontario N2L 3G1, Canada\\ 
$^{6}$Rubin Observatory Project Office, 950 N.\ Cherry Ave., Tucson, AZ 85719, USA\\ 
$^{7}$Finnish Centre of Astronomy with ESO (FINCA), Vesilinnantie 5, FI-20014 University of Turku, Finland\\ 
$^{8}$Space Physics and Astronomy Research Unit, University of Oulu, P.O. Box 3000, FI-90014 Oulun yliopisto, Finland\\ 
$^{9}$Specim, Spectral Imaging Ltd., Elektroniikkatie 13, FI-90590 Oulu, Finland\\ 
$^{10}$Steward Observatory, University of Arizona, 933 North Cherry Avenue, Tucson, AZ 85721, USA\\ 
$^{11}$Center for Data Intensive and Time Domain Astronomy, Department of Physics and Astronomy, Michigan State University, East Lansing, MI 48824, USA\\ 
$^{12}$Centre for Astrophysics and Supercomputing, Swinburne University, John Street, Hawthorn VIC 3122, Australia\\ 
$^{13}$School of Physics and Astronomy, RIT, Rochester, NY 14623, USA\\ 
$^{14}$Institute for Advanced Study, 1 Einstein Drive, Princeton, NJ 08540, USA\\ 
$^{15}$University of Tampa, 401 West Kennedy Boulevard, Tampa, FL 33606, USA\\ 
$^{16}$CCAPP and Department of Astronomy, The Ohio State University, 140 W. 18th Ave., Columbus, OH 43210, USA\\ 
$^{17}$Space Telescope Science Institute, 3700 San Martin Drive, Baltimore, MD 21218, USA\\ 
$^{18}$Department of Physics, Engineering Physics and Astronomy, Queen's University, Kingston, ON K7L 3N6, Canada \\ 
$^{19}$Instituto de Astrof\'isica de Canarias, E-38205 La Laguna, Tenerife, Spain\\ 
$^{20}$Departamento de Astrof\'isica, Universidad de La Laguna, E-38205 La Laguna, Tenerife, Spain\\ 
$^{21}$NSF’s National Optical-Infrared Astronomy Research Laboratory, 950 N. Cherry Ave., Tucson, AZ 85719, USA\\ 
$^{22}$Department of Physics \& Astronomy, University of North Carolina at Chapel Hill, 120 E.\ Cameron Ave, Chapel Hill, NC 27599, USA\\ 
$^{23}$Royal Military College of Canada, PO Box 17000, Station Forces, Kingston, Ontario, Canada K7K 7B4\\ 
$^{24}$Astronomy Department, Yale University, 52 Hillhouse Avenue, New Haven, CT 06511, USA 
}

\date{Accepted XXX. Received YYY; in original form ZZZ}

\pubyear{2022}
 
\begin{document}
\label{firstpage}
\pagerange{\pageref{firstpage}--\pageref{lastpage}}
\maketitle

\begin{abstract}
The classical globular clusters found 
in all galaxy types have 
half-light radii of $r_{\rm h} \sim$~2--4 pc,
which have been tied to formation in the dense cores of giant molecular clouds.
Some old star clusters 
have larger sizes, and it is unclear if these 
represent a fundamentally different mode of low-density star cluster formation.
We report the discovery of a rare, young ``faint fuzzy'' star cluster, 
NGC~247-SC1, on the outskirts of the low-mass spiral galaxy NGC~247 
in the nearby Sculptor group,
and measure its radial velocity using Keck spectroscopy.
We use {\it Hubble Space Telescope} imaging to measure the cluster half-light radius
of $r_{\rm h} \simeq 12$~pc
and a luminosity of $L_V \simeq 4\times10^5 \mathrm{L}_\odot$.
We produce a colour--magnitude diagram of cluster stars and compare to theoretical isochrones, finding an age of $\simeq$300~Myr, a metallicity of [$Z$/H]~$\sim -0.6$
and an inferred mass of $M_\star \simeq 9\times10^4 \mathrm{M}_\odot$.
The narrow width of blue-loop star magnitudes implies an age spread of
$\lesssim$~50 Myr, while no old red-giant branch stars
are found, so SC1 is consistent with hosting a single stellar population, modulo
several unexplained bright ``red straggler'' stars.
SC1 appears to be surrounded by tidal debris, at the end of
a $\sim$~2~kpc long stellar filament that also hosts two low-mass, low-density clusters of 
a similar age.
We explore a link between the formation of these unusual clusters and an external perturbation of their
host galaxy, illuminating a possible channel by which some clusters are born with large sizes. 
\end{abstract}





\section{Introduction}\label{sec:intro}



Old globular clusters (GCs) have been observed in a wide range of host galaxies to have fairly homogeneous properties that point to universal mechanisms or initial conditions for long-lived star cluster formation.
For example, their typical half-light radii have a limited range of $r_{\rm h} \sim$~2--4 pc across a stellar mass range of $M_\star \sim 10^5$--$10^6 \mathrm{M}_\odot$,
while they host distinctive
``multiple populations'' of stars characterized by variations in light element abundances (e.g.\ \citealt{Bastian18}). 
Young star clusters
with similar properties to GCs have been found in nearby star-forming regions 
(e.g.\ \citealt{Larsen04,Brown21}), with sizes that are thought to be linked to the densities of their progenitor giant molecular clouds
(e.g.\ \citealt{Grudic21,Grudic22}), 
and emerging evidence of multiple populations (e.g.\ \citealt{Saracino20,Li21,Cadelano22,Asad22}).
Overall, a picture has emerged
where bound star cluster formation is a continuous process from early times to the present day, albeit with 
fewer and fewer massive GCs being formed
while massive, dense gas clouds become ever rarer.

This tidy picture was disturbed by discoveries of new classes of old star clusters with more diverse properties than the classical GCs.
These novelties included ultracompact dwarfs (UCDs; \citealt{Hilker99,Drinkwater00}), 
with typically larger sizes and luminosities:
$r_{\rm h} \sim$~10--100~pc and $L_V \sim 10^7 \mathrm{L}_{V,\odot}$.
A few much fainter, large clusters ($r_{\rm h} \sim$~10--30~pc and $L_V \sim 10^4 \mathrm{L}_{V,\odot}$) 
were long known in the outer halo of the Milky Way (MW), but many more were later found in dwarf galaxies, in the halo of M31 and in massive lenticular (S0) galaxies
(e.g.\ \citealt{Brodie02,Huxor05,Peng06,Hwang11}), with nomenclatures including diffuse star clusters,
extended clusters (ECs) and faint fuzzies (FFs).
The lines between these classes were further blurred with the discovery of star clusters that filled the ``gap'' between ECs/FFs and UCDs, with large sizes and intermediate luminosities \citep{Brodie11,Forbes13}.

The relations between the different star cluster families are unclear,
but multiple formation pathways seem likely.
Many of the UCDs are now thought to be stripped galactic nuclei 
(e.g.\ \citealt{Jennings15,Ahn18,Mayes21a}).
Others may be bona fide star clusters whose large sizes are the product of mergers of smaller clusters (e.g.\ \citealt{Fellhauer02}).
Similarly, multiple mechanisms have been proposed for the origins of ECs and FFs, including
special conditions in the interstellar medium during galactic collisions \citep{Burkert05,Elmegreen08},
star cluster mergers \citep{Bruns11},
expansion in the weak tidal fields of dwarfs and galaxy haloes
\citep{Madrid12}
and even the effects of stellar-mass black holes
\citep{Gieles21}.
Overall, ECs and FFs have received far less attention than GCs and UCDs, with
little to
no observational work carried out on their early formation histories or
on the presence of multiple populations.

\begin{figure}
	\includegraphics[width=\columnwidth]{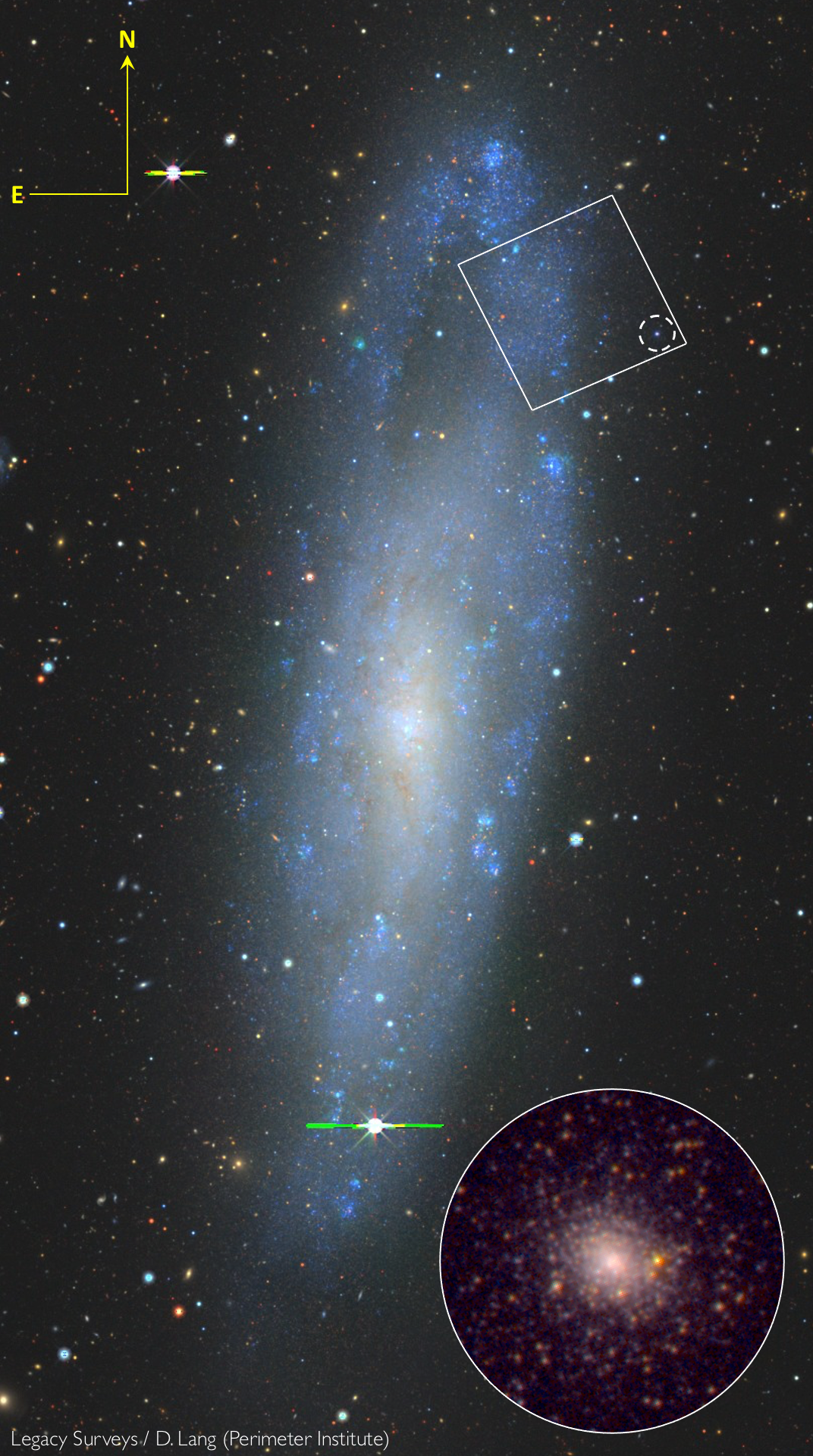} 
    \caption{Overview of the star cluster NGC~247-SC1 in context.
The larger image shows the host galaxy NGC~247 in 
$grz$ Legacy Surveys imaging\protect\footnotemark,
with a $\simeq 13\arcmin \times 24\arcmin \simeq 14\times 25$~kpc field of view.
North is up and East is left.
The quadrilateral at upper right outlines the {\it HST}/WFC3 footprint, with a dashed circle around SC1.
At lower right is a WFC3 zoom-in on SC1
($\sim 8\arcsec \sim 130$~pc diameter field of view, using F475W and F606W filters), 
which sits 8\arcmin\ (8 kpc) from the center of NGC 247 (in projection), and 
appears to be just
beyond the outer rim of the disc.
}
\label{fig:overview}
\end{figure}

Here we present an unusual star cluster, NGC~247-SC1 (hereafter SC1), associated with the disc galaxy NGC~247 in the nearby Sculptor group of galaxies
(Figure~\ref{fig:overview}).
SC1 was discovered as part of the MADCASH survey, whose goals are to study the assembly of massive dwarf galaxies through observations of their satellites, stellar haloes and GC systems (e.g.\ \citealt{Carlin16}).
From Subaru/Hyper Suprime-Cam (HSC) imaging,
SC1 stood out from other known GCs around NGC~247
by its brightness and its more extended light distribution.
After initially considering SC1 as a UCD,
we realized that the high luminosity was largely driven by its young age (discussed below), and that by mass, the cluster should be classified as an EC or FF.

The host galaxy, NGC~247, is a borderline dwarf/giant Sd galaxy with stellar mass $M_\star \simeq 3\times10^9 \mathrm{M}_\odot$, no bulge,
disc scale length $\simeq 4$~kpc,
inclination of $76^\circ$ from face-on,
disc rotation speed $\simeq 100$~km~s$^{-1}$,
halo mass of $M_{200} \sim 2\times10^{11} \mathrm{M}_\odot$
and specific star formation rate of $\sim 6\times 10^{-11}$~yr$^{-1}$
\citep{Romanowsky12,Fall18,Leroy19,Li20}.
The distance is $3.52 \pm 0.10$~Mpc \citep{Tully13},
with a corresponding linear scale of 17 pc per arcsec, and 1.0 kpc per arcmin.
All photometry in this paper is corrected for Galactic extinction using
coefficients from \citet{Schlafly11}, 
adopting the \citet{Fitzpatrick99} reddening law with $R_V = 3.1$,
as provided by the NASA/IPAC Extragalactic Database (NED):
$A_\mathrm{F225W} = 0.125$, $A_\mathrm{F275W} = 0.098$, $A_\mathrm{F390W} = 0.070$,
$A_\mathrm{F475W} = 0.058$, $A_\mathrm{F606W} = 0.045$, $A_\mathrm{F814W} = 0.028$,
$A_g = 0.057$, $A_i = 0.030$, $A_C = 0.071$, $A_M = 0.049$ and $A_{T_1} = 0.039$.

\footnotetext{\url{https://www.legacysurvey.org/}}

\begin{figure*}
	\includegraphics[width=17cm]{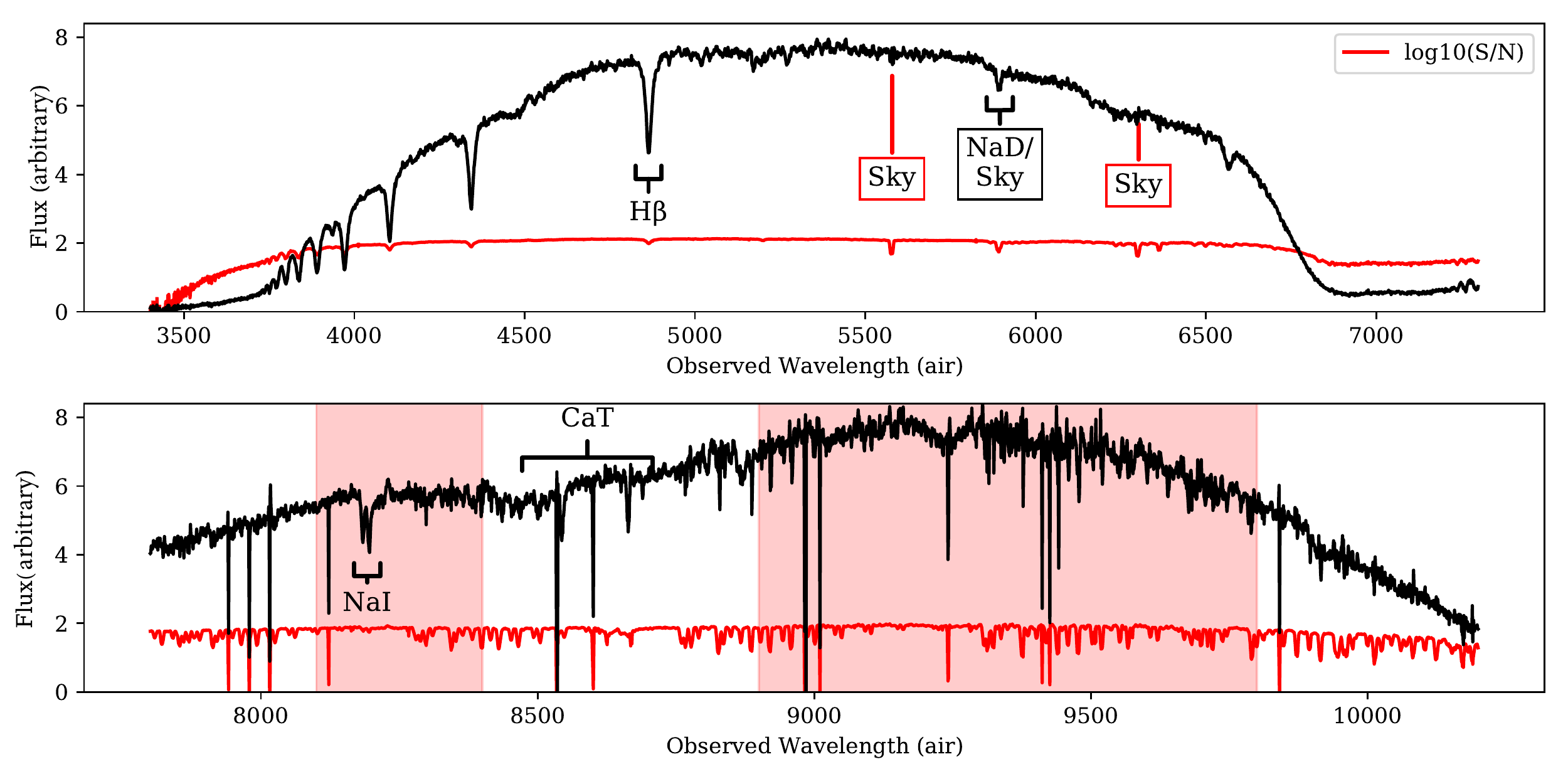}
    \caption{The coadded, telluric-corrected LRIS spectrum for SC1 (black curves),
    with the blue side in the top panel, and red side in the bottom. The S/N (in log units) as a function of wavelength is shown as red curves. We highlight important spectral features, locations of sky lines and windows of significant telluric contamination.
    Strong Balmer lines are visible which indicate a relatively young age,
    and the calcium triplet is used for the redshift estimate.
    }
    \label{fig:LRIS}
\end{figure*}

In the rest of this paper, we present spectroscopy of SC1 (Section~\ref{sec:spec}), 
photometric observations and analysis (Section~\ref{sec:phot}),
photometric results (Section~\ref{sec:photres}),
discussion (Section~\ref{sec:disc})
and conclusions (Section~\ref{sec:conc}).

\section{Spectroscopy}\label{sec:spec}

We acquired optical to near-infrared spectroscopy for SC1 using the Low Resolution Spectrometer \citep[LRIS;][]{oke1995} with the atmospheric dispersion corrector at the W.~M.\ Keck Observatory
on 27, 28 and 29 October 2016 (UT;  
program ID Y053M).
LRIS has separate blue and red channels. On the blue side, 
we used the 300/5000 grism and on the red side we used the 600/10000 grating,
with a spectral full-width-at-half-maximum (FWHM) resolution that
is strongly wavelength-dependent,
from $\sim 150$ to 300 km~s$^{-1}$ on the blue side, to $\sim 45$ km~s$^{-1}$ on the red side.
With this instrument setup, the D680 dichroic, 
and a long slit with 1\arcsec\ width and 3\arcmin\ length oriented East--West,
we achieved $\sim$~3500--7000~\AA\ wavelength coverage on the blue side and  $\sim$~7800--10200~\AA\ on the red.
The total exposure time was about 2.5~hours,
and the airmass ranged from $\sim$ 1.4 to 1.8.

We used the open source Python Spectroscopic Data Reduction Pipeline version 1.4.0 \citep[\texttt{PypeIt;}][]{pypeit_reference} to do the basic data reduction.
The bias level was estimated from the overscan region and subtracted from the raw data.
We obtained internal flat field images at the beginning of each night.
While we did not flat field the data for SC1, the flat field images were used by \texttt{PypeIt} to trace the edges of the slit of the detector and automatically detect the object in the slit.
\texttt{PypeIt} performs a 2D BSpline sky subtraction \citep[e.g.][]{kelson2002} across the entire slit.
The 1D spectra were automatically extracted using an algorithm \citep[e.g.][]{horne1986}.

\texttt{PypeIt} wavelength calibrates on the 1D spectra by calibrating on arc lamp spectra.
We obtained arc frames at the beginning of each night. For wavelength calibration on the blue side we observed the Hg, Cd and Zn arc lamps; on the red side we observed the Ne, Ar, Kr and Xe arc lamps. 
We further optimized the wavelength solution by correcting the errors introduced by the varying flexure of the instrument. 
For the red arm, we computed how offset in wavelength the observed sky emission lines were from those in a model sky spectrum. 
Due to the paucity of sky lines in the region covered by the blue arm, we cross-correlated regions spanning 250\AA\ over the observed spectrum with a template simple stellar population to compute offsets in each region.
The wavelength arrays of each arm were corrected using linear functions fitted to the respective offsets as a function of wavelength.
It is possible there are still residual flexure errors at the level of $\sim$~10~km~s$^{-1}$.

We did not flux calibrate the 1D spectra. 
The individual exposures were coadded by weighting each exposure by the inverse variance at each pixel.
The coadding also cleaned the final spectrum of cosmic rays.
For the telluric correction, we followed the methodology of \citet{vanDokkum12}
with slight modifications, performing the 
correction outside of \texttt{PypeIt} by iterating through a grid of telluric models produced from the Line-By-Line Radiative Transfer Model 
\citep{clough2005, gullikson2014}, and scaling each template to minimize the difference between it and the observed flux over the region 9320--9380~\AA. 

The LRIS slit captured not only the light from SC1 but also from a bright red star
(hereafter `St1') that is projected 1\farcs0 to the West from the centre of SC1
(see zoom-in image in Figure~\ref{fig:overview}).
This star will be discussed in more detail later in the paper. It
is well within the total extent of SC1, which means that there is 
considerable overlap between the two spectra on the red side
(the blue side spectrum appears to be completely dominated by SC1).
Even so, with seeing of $\sim 0\farcs9$, 
and the extended nature of SC1, it is possible to separate the spectra in
an approximate fashion by simply extracting
East and West halves of the spectral trace (each with a width of 8-pixels or 1\farcs0).

We show the final spectrum of SC1 in Figure~\ref{fig:LRIS} (black curves) 
with the blue-side in the top panel and the red-side in the bottom panel,
and with some of the prominent absorption lines highlighted.
We also show the wavelength-dependent signal-to-noise (red; S/N; red curves) in each panel to demonstrate the high-S/N achieved ($\sim 100$~\AA$^{-1}$).
The Balmer lines are very strong, indicating that the light is dominated by relatively young main sequence stars, with ages somewhere in the range $\sim$~0.1--1~Gyr.
However, no emission lines are seen in the spectrum.

We use the code \texttt{Prospector} \citep{johnson2021} to determine the redshift of SC1 using the calcium triplet (CaT) region. 
We masked out nonphysical artefacts 
over the spectral ranges 8535--8540~\AA\ and 8601--8606~\AA.
\texttt{Prospector} is a code for inference of physical parameters from spectroscopic data via MCMC sampling of the posterior probability distributions. 
To obtain an estimate of the posterior redshift distribution for SC1, we use a grid of theoretical stellar atmosphere models. 
At each MCMC step the spectrum is shifted in velocity and the likelihood of the data given the redshift and the smoothing needed to match the grid to the data is calculated.

The heliocentric-corrected recession velocity of SC1 is 
$112\pm5$~km~s$^{-1}$,
confirming that it is not a background galaxy (its fuzziness in the HSC imaging already ruled it out as a foreground star).
This velocity also strongly supports an association with NGC~247 as the host galaxy,
which has an overall recession velocity of $\sim$~150--160~km~s$^{-1}$ (NED).
Spectroscopic analysis of the stellar population is planned for a follow-up paper.
The same procedures with St1 return a velocity of 
$130 \pm 5$~km~s$^{-1}$,
suggesting it may be associated with SC1.
We note that because the two spectra were not completely de-blended,
the true velocity difference may be slightly larger.
There may also be systematic errors remaining in the wavelength calibration at the level of $\sim$~10 and $\sim$~20 km~s$^{-1}$ for relative and absolute velocities, respectively.
The implications of these velocities will be considered in more detail in Sections~\ref{sec:strag} and \ref{sec:filament}.

\section{Photometric observations and analysis}\label{sec:phot}

\subsection{Subaru Hyper Suprime-Cam observations}\label{sec:hsc}

We used the 
HSC imager on the 8.2m Subaru telescope \citep{Miyazaki12}
to observe a single pointing centred on NGC~247 on 15 Oct 2015. The observations totalled 5040s (16 exposures of 315s each) in $g$-band (known as ``HSC-G'' at Subaru) and 2565s (9$\times$285~s exposures) in $i$-band (``HSC-I'').
The repeated exposures were dithered translationally and rotationally to fill in chip gaps and to enable cosmic-ray removal. We also took a single 30~s image in each filter to extend the dynamic range of our observations to a brighter saturation limit. The data were all obtained during photometric conditions, with seeing between 0\farcs55 and 0\farcs7 for all frames. We processed the raw data with a development version of the Rubin/LSST software pipelines, which have also been applied to HSC data (e.g.\ \citealt{Bosch18, Bosch19, Aihara18a, Aihara18b, Aihara19}), including removal of instrument signal, image coaddition, source detection and measurement. The results presented in this work are derived from point spread function (PSF) photometry, astrometrically calibrated to Gaia DR2 \citep{Gaia18} and photometrically calibrated to the PanSTARRS-1 photometric system 
\citep{Schlafly12, Tonry12, Magnier13}.

In order to extract photometry more optimally in the crowded environs of SC1 (and near the main body of NGC~247 itself), we ran additional PSF fitting photometry on the images that were processed by the LSST pipeline using {\tt DAOPHOT}, {\tt ALLSTAR} and {\tt ALLFRAME} 
\citep{Stetson87,Stetson92,Stetson94}. This extra measurement step was performed only on a single 4k by 4k ``patch'' from the HSC data; crowding from stars in the outer disc of NGC~247 becomes too extreme to extract measurements just to the east of this patch. We ran two iterations of {\tt ALLSTAR}, first on the original science images, then again after subtracting sources measured on the first pass to recover faint stars missed on the first iteration. We then performed forced photometry using {\tt ALLFRAME} at the locations of all sources detected in either $g$ or $i$ bands in order to improve our photometric depth. Only sources with final {\tt ALLFRAME} measurements in both $g$ and $i$ are kept in the final catalog. We utilize stars in common between the {\tt ALLFRAME} and LSST photometric catalogs to bootstrap the {\tt ALLFRAME} photometry onto the LSST calibration. The final {\tt ALLFRAME} catalog shows very good agreement with the LSST catalog for shared sources, with a residual standard deviation of about 0.05 mag at $g=25.5$ mag. The advantage of the {\tt ALLFRAME} catalog is that it is much more complete for regions of high stellar density, allowing us to extend our star map further to the east.

To select stars, we make use 
of the {\tt DAOPHOT} ``sharp'' parameter
and the $\chi$ statistic,
both of which measure deviations from the PSF model
(sharp $\simeq 0$ and $\chi \simeq 1$ are good results for a point source).
At the faint magnitudes of interest here ($i \gtrsim 24.5$), 
we find no clear demarcation in sharp and $\chi$ between the stellar locus and other objects.
Therefore to help guide the selection,
we use a cross-matched catalogue from {\it HST}, which has much better discrimination
between stars and extended objects (see next Section).
We adopt cuts based on the $i$-band imaging (with the best seeing)
of $-0.5 < \mathrm{sharp} < 0.8$ and $0.8 < \chi < 1.4$,
which will reduce but not completely eliminate extended objects and blends of stars.
The results using HSC star-counts later in the paper (Section~\ref{sec:filament}) 
are not sensitive to the boundaries of these cuts.

\subsection{{\it Hubble Space Telescope} observations}\label{sec:hst}

NGC~247-SC1 was observed with {\it HST} as part of Program ID 14748
(PI: A.~Romanowsky), on 21 and 24 June 2017, using the Wide Field Camera 3 imager with the UVIS channel.
Images were taken through the F475W and F606W filters with total exposure times of 6956~s and 7274~s, respectively, split into six exposures per filter 
(plus two shorter exposures, see below).  
These filters were optimised for constraining the age distribution of stars in SC1.
A three-point subpixel dither pattern was used (one point per orbit) with
a two-point line sub-pattern (each orbit split into two exposures).
The target cluster was placed near a read-out amplifier in a corner of the field in order to 
reduce charge transfer efficiency (CTE) losses;
the opposite side of the field includes a high-density region of the host-galaxy disc
(see Figure~\ref{fig:overview}).

Shorter exposures were also taken with the F225W, F275W, F390W and F814W filters, 
with exposure times of $\sim$~100--300~s each, for increased wavelength range for
stellar population diagnostics using integrated light.
Additional short exposures (60~s) were taken in F475W and F606W.
All of the short exposures used a two-point dither pattern and post-flash illumination of 12 electrons per pixel to help with CTE losses.

The images have a pixel scale of 0\farcs04 (0.7~pc) and a field of view of 2\farcm7$\times$2\farcm7 (2.8$\times$2.8~kpc).
We downloaded {\tt drc} (drizzled and CTE-corrected) images from the Mikulski Archive for Space Telescopes (MAST) for the photometry.
We applied extinction corrections to the photometry 
as given in Section~\ref{sec:intro}.

In addition to these observations, we generate simulated data-sets
to be analyzed in parallel, in order to assess the reliability
of the results and to carry out fair comparisons to models
(see \citealt[hereafter L11]{Larsen11}).
We generated artificial clusters resembling SC1 and inserted them
into the F475W and F606W images, using the {\tt MKSYNTH} task in the {\tt BAOLAB} package
\citep{Larsen99}.
The procedure here is to draw artificial stars
at random from a cluster resembling SC1 in its density profile and 
its stellar population.
{\tt MKSYNTH} models the artificial stars by considering the PSFs (derived in Section~\ref{sec:ptphot}) as probability density functions, centred on the coordinates of each star, from which individual counts are drawn and added to the image.  The artificial clusters were placed in relatively empty areas of the image,
with a default position $\sim$20\arcsec\ to the ENE from SC1.
This procedure allows us to include the effects of contamination in the analysis --
both from the host galaxy and from foreground stars.

\begin{figure}
	\includegraphics[width=\columnwidth]{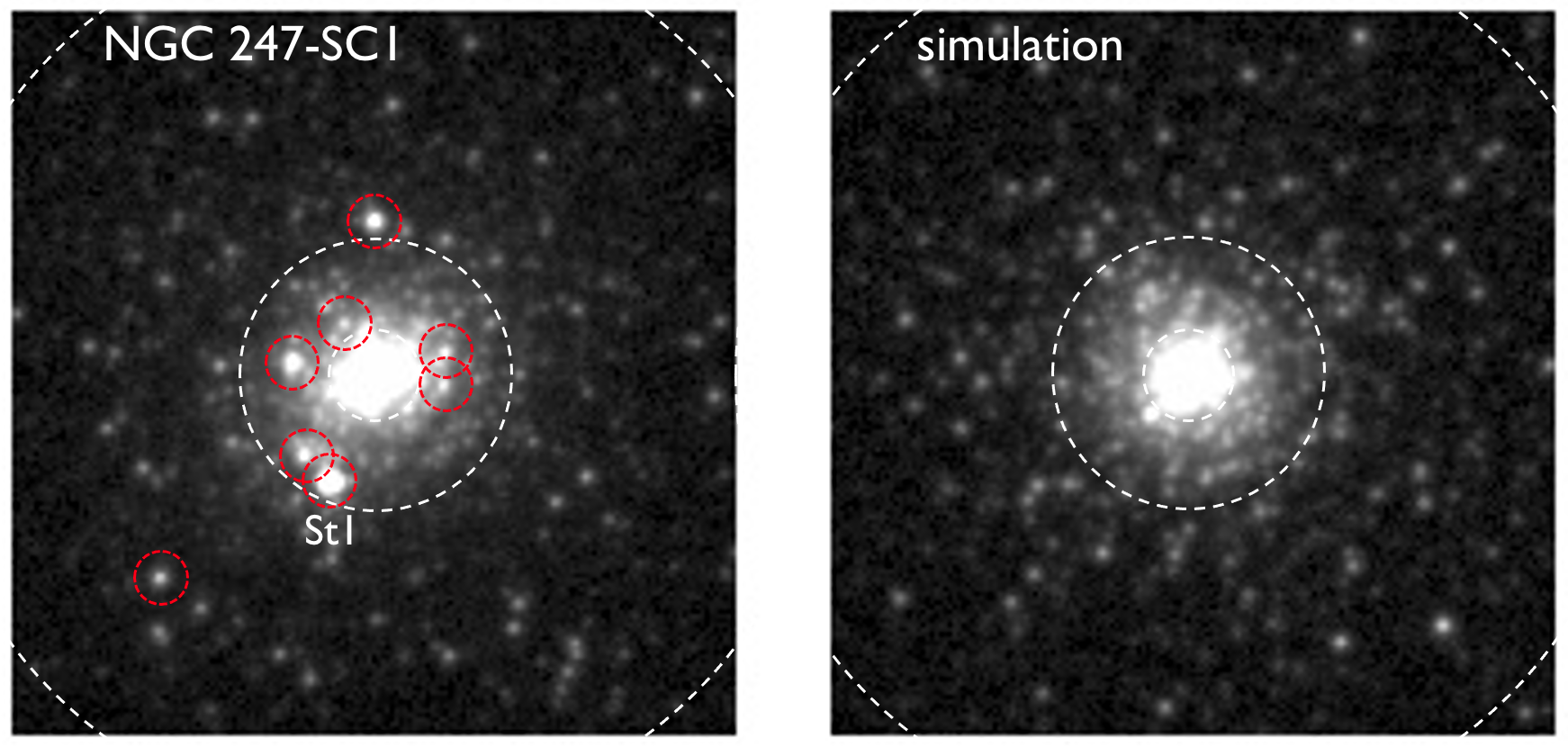}
    \caption{Real ({\it left}) and simulated ({\it right}) WFC3 images of SC1 in the F606W filter.
    The field of view is 160$\times$160 pixels = $6\farcs4\times6\farcs4$ = $110\times110$~pc.
    Dashed white circles (radii of 10, 30, 100 pixels) mark the boundaries of the two annuli used
    for the colour--magnitude diagrams.
    Smaller red dotted circles in the left panel mark ``red straggler'' stars,
    with the very bright non-member star St1 labelled;
    one additional red straggler is outside the field of view. 
    }
    \label{fig:syntimg}
\end{figure}

For each star, a mass was drawn randomly from a 
Kroupa initial mass function (IMF), and F475W and F606W magnitudes were
assigned by interpolation in a PARSEC isochrone \citep{Marigo17}\footnote{\url{http://stev.oapd.inaf.it/cgi-bin/cmd}}
with $Z=0.004$, i.e. [$Z$/H]~$=-0.6$, 
and an age of 316 Myr ($\log t = 8.5$; these values
will motivated in the Section~\ref{sec:cmd}).
The number of stars was
adjusted to reproduce the total integrated F606W magnitude of SC1
within a 200~pixel aperture.  
The real and simulated F606W
images are shown in Figure~\ref{fig:syntimg}. Note that the real
image contains several bright stars, which are not present 
in the simulated image. These will be discussed further below.
Photometry was then carried out on the simulated images in the same
way as for the real images,
and used in the next Sections.

\subsection{Integrated light surface photometry}\label{sec:ILphot}

We measure basic photometric parameters of the object from {\it HST} using 
cumulative aperture photometry,
and report them in Table~\ref{tab:clust_props}.
This approach has the advantages of being
simple and non-parametric, although there are also complications in 
determining the total amount of light.
A correction for the ``background'' level is required, which is non-trivial
given the likely presence of extra-tidal debris (see Section~\ref{sec:filament}).
Also, the radial light profile of the cluster appears to fall off relatively slowly
(see below), and extrapolating it to infinite radius does not make physical sense.
We instead define a radius of 200 pixels = 8\arcsec~$\simeq$~140~pc as the edge
of the cluster, 
which we will later see corresponds to the approximate tidal radius,
and measure the sky levels from just outside this radius (200--250~pixels).
We also make use of simulated cluster images (Section~\ref{sec:hst})
to test the accuracy of our results.

\begin{table}
	\centering
	\caption{Properties of star cluster NGC~247-SC1.
	All photometry is corrected for Galactic extinction.
	These include the right ascension, declination, magnitudes and colours in the {\it HST}  
	filters, 
	the $V$-band absolute magnitude, the stellar mass, the half-light radius,
	the mean stellar surface density within the half-light radius,
	the axis ratio and position angle, the adopted distance and the recession velocity.}
	\label{tab:clust_props}
	\begin{tabular}{lcc} 
		\hline
		property & value & units \\
		\hline
		R.A. & 11.71154 & deg J2000 \\ 
		Decl. & $-$20.65142 & deg J2000 \\ 
        $\mathrm{(F606W)}_0$ & $18.35 \pm 0.02$ & Vega mag \\ 
        $\mathrm{(F225W-F606W)}_0$ & $0.40 \pm 0.01$  & Vega mag \\ 
        $\mathrm{(F275W-F606W)}_0$ & $0.32 \pm 0.03$  & Vega mag \\ 
        $\mathrm{(F390W-F606W)}_0$ & $0.33 \pm 0.02$  & Vega mag \\ 
        $\mathrm{(F475W-F606W)}_0$ & $0.22 \pm 0.01$  & Vega mag \\ 
        $\mathrm{(F606W-F814W)}_0$ & $0.33 \pm 0.01$  & Vega mag \\ 
        $M_{V,0}$ & $-9.28 \pm 0.06$ & Vega mag \\
        $M_\star$ & $(8.9 \pm 0.5)\times 10^4$ & ${\rm M}_\odot$ \\
        $r_{\rm h}$ & $11.7 \pm 0.5$ & pc \\
        $\Sigma_{\star, \rm h}$ & $103\pm10$ & $\mathrm{M}_\odot$~pc$^{-2}$ \\ 
        $b/a$ & $0.79 \pm 0.02$ & \\ 
        P.A. & 54 & deg \\ 
        distance & $3.52 \pm 0.10$ & Mpc \\
        $v$ & $112\pm5$ & km~s$^{-1}$ \\ 
		\hline
	\end{tabular}
\end{table}

We use this procedure to derive total magnitudes of $\mathrm{(F475W)}_0 = 18.60 \pm 0.01$ 
and $\mathrm{(F606W)}_0 = 18.35 \pm 0.02$, 
where the uncertainties are derived from the uncertain background levels.
Note that for comparison to any idealized stellar population model, the expected magnitude
will be uncertain at the $\pm 0.04$~mag level
owing to stochasticity (as we have found from simulating the observations).
We interpolate between the two filters to the Johnson $V$-band to
find $V \simeq 18.45 \pm 0.02$, with an implied $M_{V,0} = -9.28 \pm 0.06$
and $L_V \simeq (4.3 \pm 0.3) \times10^5 \mathrm{L}_\odot$
(taking into account the distance uncertainty).

These total magnitudes should not be used for precise measurement of colour,
which we have found can exhibit large, erroneous excursions from the larger apertures,
probably owing to the effects of  
background contamination.
Instead, we use an aperture of $\sim$~30 pixels for the colour measurement,
and find $\mathrm{(F475W-F606W)_0} = 0.22 \pm 0.01$.
We also use this method to measure colours using the other filters with shorter exposures,
with results reported in Table~\ref{tab:clust_props}.

The half-light radius is derived by using the cumulative photometry within 200~pixels,
and is $r_{\rm h} = 0\farcs69 \pm 0\farcs03 = 11.7 \pm 0.5$~pc.
Here we adopt the average measurement from the F475W and F606W bands,
and the difference between these bands as the uncertainty
(which is consistent with stochastic differences in size measurements of a simulated cluster).

Note that the PSF is not taken into account
in this approach, but the impact on the $r_{\rm h}$ measurement should be very minor.
Also, the bright red star St1 contributes 2\% and 6\% of the light in F475W and F606W, respectively, and since it is apparently not a member of SC1 (see Section~\ref{sec:spec}), 
we subtracted its flux before carrying out the procedures above
(which affects the F606W magnitude and the overall colour at the $\sim$~0.05~mag level).

We also used the aperture flux measurements to generate
a F475W$-$F606W colour profile of SC1
out to a radius of $\sim$~1\arcsec, 
where our cluster simulation suggests variations at the
$\sim$~0.05 mag level or more would be detectable
above the stochastic background of the individual stars.
We found no indication of a colour gradient at this level,
and not in the other, shorter-exposure colours either.

\begin{figure}
{\centering
	\includegraphics[width=3.5cm]{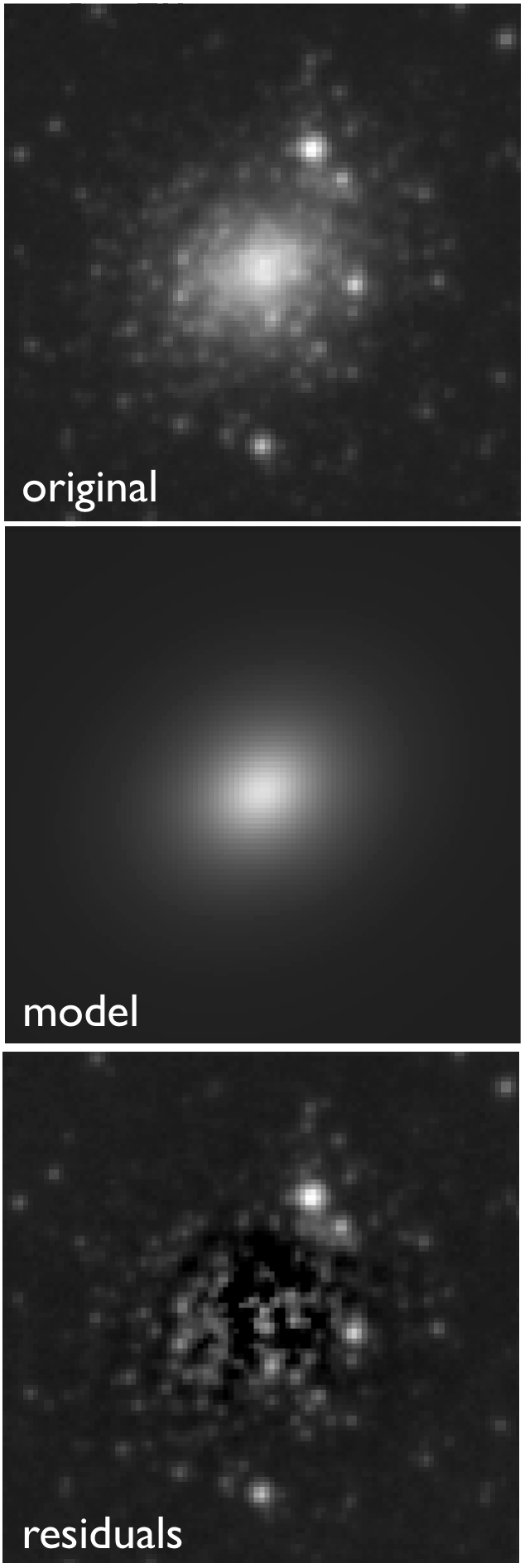} 
    \caption{Modelling of the two-dimensional surface brightness profile of SC1 (F606W image)
    using {\tt ISHAPE}. On the top is the original image, in the middle is the PSF-convolved model,
    and on the bottom are the residuals.
    The brighter point sources were essentially masked from the model fit by a weighting map.
    }
}
    \label{fig:ishape}
\end{figure}

We next use a complementary approach, the {\tt ISHAPE} task for
PSF-convolved and parameterized surface brightness modelling \citep{Larsen99}.
{\tt ISHAPE} is designed to model marginally resolved stellar systems,
and there can be complications for
a case like SC1 which is highly resolved into stars.
Keeping this caveat in mind,
we find the surface brightness to be well approximated by a Moffat (EFF) profile,
\begin{equation}
  I(r) = I_0 \left[ 1 + (r/r_c)^2 \right]^{-\eta} ,
 \label{eq:EFF}
\end{equation}
with an envelope slope index of $\eta=1.33$ and a major-axis
FWHM of 9.3 pixels, 
corresponding to $r_{\rm h} = 0\farcs54$ (circularized)
or 9.2~pc at the distance of NGC~247, with variations at the
$\sim$~5\% level depending on the filter used and the fitting radius.
Figure~\ref{fig:ishape} shows the original image, the smooth 
two-dimensional model and the residuals after model
subtraction, where a fair number of resolved stars are visible
that were effectively masked out in the fitting through
a weighting function.
Alternatively removing the weighting gives a larger radius:
$r_{\rm h} = 0\farcs63$ or 10.8~pc.
The latter value is consistent with the aperture-based measurement,
which we will adopt as our best estimate given the
resolved nature of the system
(although we take the axis ratio and position angle values
from the {\tt ISHAPE} fits, listed in Table~\ref{tab:clust_props}).
We note also that {\tt ISHAPE}-based estimates from
the HSC imaging also gave $r_{\rm h} \simeq 10$~pc.

We also measure the integrated-light photometry of SC1 from HSC, using
aperture photometry within a radius of 3\farcs6, corrected by a background level within an annulus of equal area at larger radii.
To capture the variations in the background, we use random quarter-circle annuli at radii between 2 and 4 times the aperture radius.
We find $g_0 \geq 18.62 \pm 0.03$ AB mag and $(g-i)_0 \leq 0.37 \pm 0.04$ AB mag,
where the uncertainties reflect both photometric noise and background variations,
and the inequalities reflect the presence of the red star St1 which has not been
de-blended in the photometry.

There is furthermore ground-based photometry available
in Washington filters $CMT_1$ from the Mosaic~II imager on the Cerro Tololo Inter-American Observatory (CTIO) 4~m telescope.  
This is described in \citet{Olsen04}, where SC1 was not reported in the catalogue of candidate GCs owing to its colour being too blue for an old stellar population.
It has a blue magnitude of $C_0 \geq 18.72$,
and colours of $(C-M)_0 \leq 0.11$ and $(C-T_1)_0 \leq 0.53$
(again with inequalities owing to the St1 blend).

\subsection{Point source photometry}\label{sec:ptphot}

We used the {\tt ALLFRAME} package \citep{Stetson94} to carry out
PSF-fitting photometry on the drizzle-coadded images, following a similar
procedure to that described in L11.
First, the {\tt FIND} task in {\tt DAOPHOT} was used to detect point sources in the
F606W images, and a PSF was then generated for each image
with the PSF task, using 20 isolated, relatively bright stars. A first pass of
PSF-fitting photometry was then obtained by letting {\tt ALLFRAME} 
measure stars simultaneously in the F475W and F606W images. In a second iteration,
improved PSFs were constructed from images in which all stars except the
PSF stars had been subtracted, and the FIND task was applied to the
star-subtracted images to detect any sources missed in the first pass.
The combined source lists were then used as input to a second pass of
{\tt ALLFRAME}.

The {\tt ALLFRAME} photometry was calibrated to standard VEGAMAG magnitudes by carrying out aperture photometry on the PSF stars in an $r=5$ pixels ($0\farcs2$) aperture.
The mean difference between the {\tt ALLFRAME} and aperture magnitudes was then added back to the {\tt ALLFRAME} magnitudes with an additional correction of $-0.18$ mag to account for the encircled flux within the reference aperture\footnote{\url{https://www.stsci.edu/hst/instrumentation/wfc3/data-analysis/photometric-calibration/uvis-encircled-energy}}. 
Photometric zero-points were adopted from the WFC3 pages at STScI\footnote{\url{https://www.stsci.edu/hst/instrumentation/wfc3/data-analysis/photometric-calibration/uvis-photometric-calibration}}.
The photometry was also corrected for extinction,
and is complete to $\sim$~29~mag in both bands.

\begin{figure*}
\includegraphics[width=18cm]{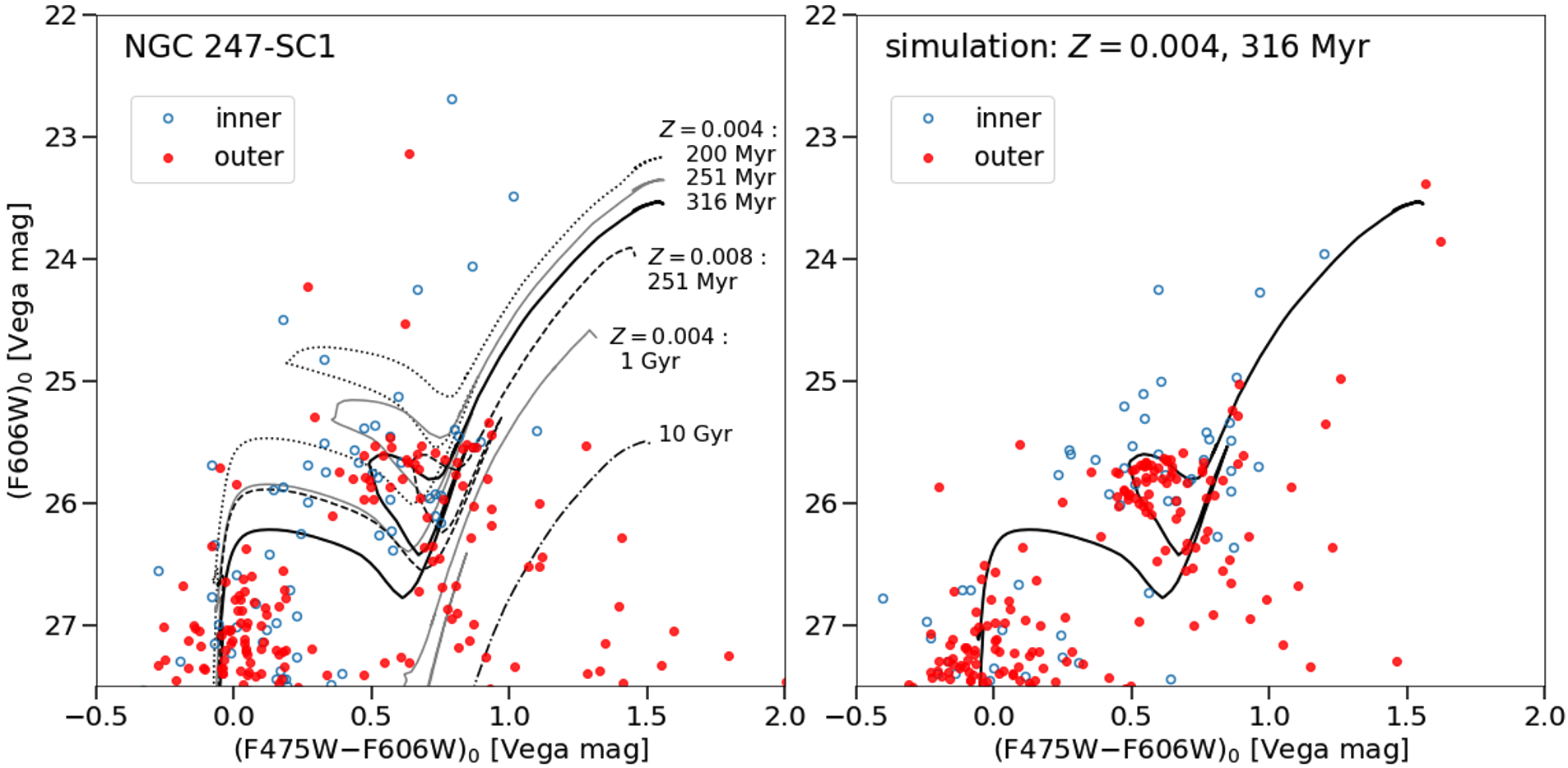}
\caption{Colour--magnitude diagram of point sources around SC1,
with left panel showing the actual observations (extinction-corrected), and right panel showing simulated data.
Coloured circles correspond to stars selected from two different annuli, at distances of 10--30 pixels (open blue) and 30--100 pixels (filled red).
Isochrones are shown at a range of ages from 200 Myr to 10 Gyr, as
labelled in the diagram (grey and black curves);
these all have $Z=0.004$ except for one case of $Z=0.008$.
The observations match up well overall with the 316~Myr isochrone
around both the main-sequence turn-off and the blue-loop that traces
the stage of core helium burning.
The simulated dataset on the right 
uses this age and metallicity solution, with results that
generally reproduce the observations on the left,
both in the distributions and the numbers of the stars.
On the other hand, the observations are missing AGB stars 
(bright and very red)
predicted by the simulation, 
and have an excess of
bright stars of intermediate colours that are difficult to explain
(see main text).
}\label{fig:CMD}
\end{figure*}

\section{Photometric results}\label{sec:photres}

The overall colour of SC1 from the WFC3 imaging, $(\mathrm{F475W}-\mathrm{F606W})_0 = 0.22 \pm 0.01$ (Vega),
corresponds to $(B-V)_0 = 0.34 \pm 0.01$, 
using the {\it HST} photometric conversion tool\footnote{\url{https://colortool.stsci.edu/uvis-filter-transformations/}}
\citep{Sahu14}.
This colour is much bluer than observed for classical old, metal-poor GCs (e.g.\ \citealt{Reed88}),
supporting a young age as implied by the spectroscopy.
Similarly, the CTIO Washington colour $(C-T_1)_0 \leq 0.53$ 
also implies an upper-limit on the age of $\sim$~0.5--1 Gyr, depending on the
metallicity (e.g.\ Fig.~14 of \citealt{Richtler12}). 
We return to more detailed consideration of the integrated colour implications later in this Section.

The remainder of this Section is structured as follows.  
Section~\ref{sec:cmd} provides the core results of this paper, with
analysis of the color--magnitude diagram (CMD) of SC1
to estimate age and metallicity, while leveraging the CMD of the simulated cluster for reference.
Section~\ref{sec:intcol} compares integrated colours of SC1 to model predictions, and
Section~\ref{sec:dage} tests for an age spread.
Section~\ref{sec:strag} explores some unexpected bright red straggler stars, while
Section~\ref{sec:mass} estimates the cluster mass.

\subsection{Color--magnitude diagram analysis}\label{sec:cmd}

The CMD of point sources around SC1 is shown in
the left panel of Figure~\ref{fig:CMD}, in two different annuli
(see dashed circles in Figure~\ref{fig:syntimg}):
10--30 pixels = 0\farcs4--1\farcs2 $\simeq$~7--20~pc $\simeq$~0.7--2.2~$r_{\rm h}$
(blue points; at smaller radii than these, crowding becomes too severe);
and 30--100 pixels = 1\farcs2--4\farcs0 $\simeq$~20--70~pc $\simeq$~2.2--7.4~$r_{\rm h}$
(red points).
Here, and for the remainder of the analysis, we adopt a cut on the photometric catalogues of $\chi < 1.5$ in F606W. 
This cut greatly reduces the scatter in the points from the crowded inner annulus,
and has little effect in the outer annulus.
We have also examined cuts using the sharp parameter, which generally rejects the same objects as $\chi$ but is less restrictive, while for the brightest stars it tends to be too restrictive.  Thus we have adopted cuts in $\chi$ only.

The right panel shows the same diagram, but using a simulated dataset
(with $Z = 0.004$ and 316~Myr age; Section~\ref{sec:hst}),
and with the same $\chi$ cut
(there are 25 contaminants in this control field, mostly with (F606W)$_0 > 26.7$).
It is apparent that the photometric depth 
of the dataset is good enough to study the features of interest in the CMD,
and the limiting factors will be crowding
and small-number statistics of the high-mass stars.
The crowding effects, even with the use of the $\chi$ cut, 
can be appreciated by the greater scatter
in the blue than the red points in the right-hand panel, 
since these are all drawn from the same isochrone.
In the remainder of the discussion, we will refer to the F475W and F606W 
filters as $g$ and $V$ for brevity.

PARSEC isochrone curves are also shown in Figure~\ref{fig:CMD} for reference,
with $Z = 0.004$ as in the simulation,
and five steps in age from 200~Myr to 10 Gyr, as labelled in the diagram;
one additional isochrone is shown for $Z=0.008$ and 251~Myr (dashed black curve).
The observations show a vertical ridge in the CMD of very blue stars 
with $g-V \sim 0.0$ 
extending up to $V \sim 26$ ($M_V \sim -2$), 
which corresponds to the main sequence.
At slightly brighter magnitudes and redder colours ($g-V \sim$~0.4--0.9),
there is a horizontal clump of stars in Figure~\ref{fig:CMD}
that corresponds to the blue loop (core helium burning supergiant phase) 
and matches up well with the 316~Myr isochrone,
in both the zeropoints and spreads of colour and magnitude.
Alternative solutions would be inconsistent with the data,
as illustrated by the isochrone curves.
For example, the blue loop for younger ages would be too bright and extended in colour.
Adopting a higher metallicity with a younger age could match the blue-loop zero-point
but would have insufficient colour spread.
Similarly, a lower metallicity with older age would have too large of a colour spread.

\begin{figure*}
    \includegraphics[width=17.5cm]{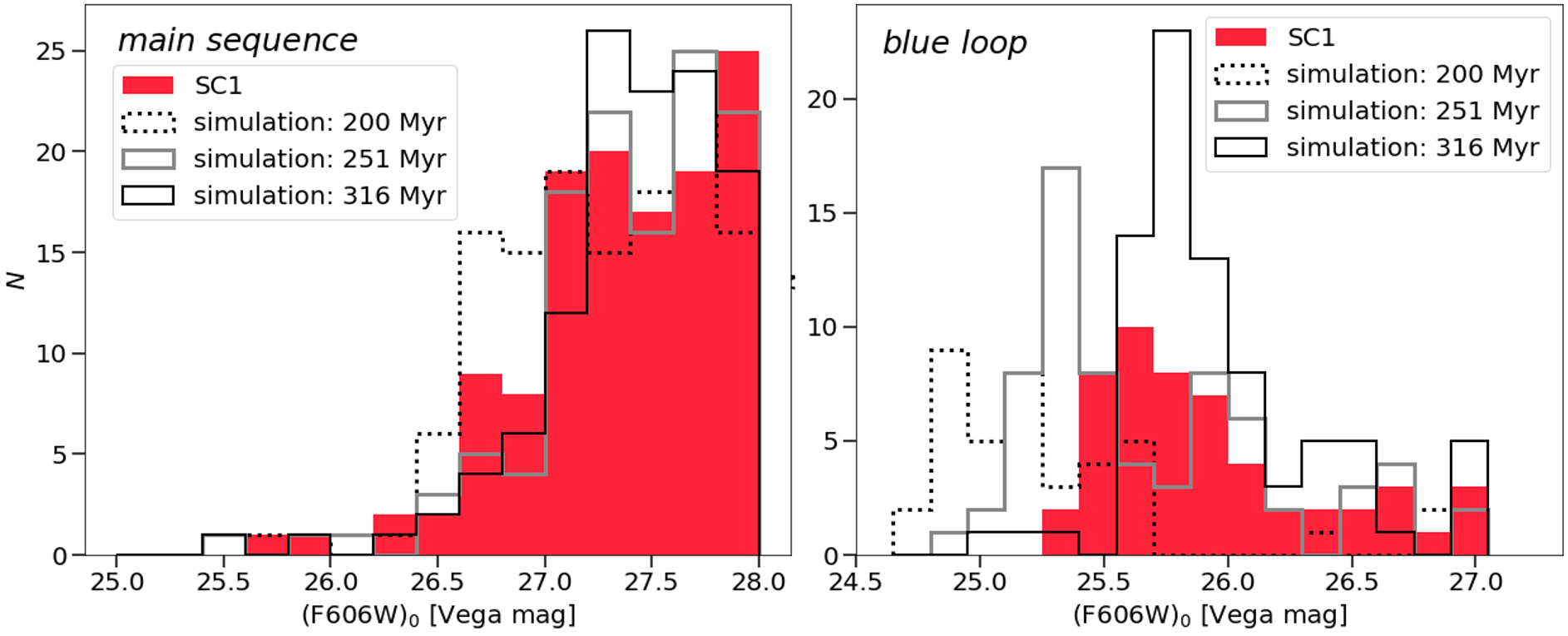} 
    \caption{Distribution of star magnitudes in the outer annulus
    on the main sequence {\it (left panel)}
    and from the blue loop {\it (right panel)}.
    The solid red histograms show the real observations of SC1,
    while the open histograms show simulated cluster observations,
    with different line shadings and styles indicating different ages 
    as in the legend ([$Z$/H] = 0.004 in all cases).
    Contamination is included in all of the histograms, but should be minimal
    over these small areas.
    Both the main sequence and the blue-loop luminosity distributions in SC1
    are consistent with an age of $\sim$~300~Myr,
    although the blue loop is more age-sensitive.
    The width of the blue loop luminosity peak ($\sigma_V \simeq 0.20$~mag)
    is consistent with the
    simulated peak, indicating no age spread at the level of $\sim$~50~Myr.
    }
    \label{fig:agehist}
\end{figure*}

Besides comparing to the idealized isochrone curves, we also compare
to the simulated cluster corresponding to the best-matching isochrone (Section~\ref{sec:hst}),
with results shown in the right panel of Figure~\ref{fig:CMD}.
The observed and simulated distributions are remarkably similar overall,
with some differences in the brightest stars that will be discussed below.
Furthermore, the absolute numbers of stars are in fair agreement,
which will be discussed further below and is effectively an 
independent test of the solution since the simulation is normalized
to reproduce the total $V$-band flux of SC1 and not the numbers of stars.
Note that this comparison also provides a
nontrivial test of the underlying stellar evolution models, since there have been challenges in the past with reproducing 
the distributions of stars along the blue loop in clusters,
as well as the ``Blue Hertzsprung Gap'' between the main sequence and the supergiants (see L11 and references therein).

To quantify the isochrone constraints further, we generate
luminosity functions of stars selected from restricted regions of the CMD.
The main sequence is isolated using
a colour range 
$-0.5 < (g-V)_0 < 0.2$,
and the observed distribution is compared in Figure~\ref{fig:agehist}
(left panel) to simulated observations over a range of ages. 
There is a great deal of stochasticity
for the very brightest magnitudes ($V_0 \sim$~25--26), owing in part to
scatter of blue-loop stars into the CMD selection window.
A more reliable metric is the onset of the rise in the luminosity function
which begins sharply at $V_0 \simeq 26.5$ for an age of 200~Myr
and more gradually at $V_0 \simeq 26.7$ for 251 and 316~Myr.
The observations of SC1 are reasonably consistent with these two older ages,
but not with the younger age.
The expected numbers of main sequence stars vary weakly with age and are in
excellent agreement with the observed number.

Next considering the blue loop,
we select stars from a box-region on the CMD
that encompasses the colour range of 
$0.2 < (g-V)_0 < 0.95$
and the magnitude range of
$V =$~24.7--27.0.
We show the resulting luminosity function
in the right panel of Figure~\ref{fig:agehist},
where the peak in the observed histogram is at $V \simeq 25.68^{+0.09}_{-0.03}$
(uncertainty from bootstrap resimulation). 
We compare to the simulated histograms of blue-loop star magnitudes, which show a strong
age dependence of $\simeq 0.5$~mag per 0.1 dex in age,
with peak magnitudes (including simulated errors) of $V \simeq 25.33 \pm 0.02$ and $25.79 \pm 0.03$, 
respectively, for ages of 251 and 316 Myr.
We can thereby formally constrain the age to be $\simeq 300 \pm 10$~Myr, 
although the systematic uncertainties are actually much larger than this.
These include the distance at the $\sim 0.1$ mag level and the choice of isochrones
(MIST predicts the stars to be fainter by $\sim$0.05~mag than in PARSEC; \citealt{Choi16}).
Arguably the most important factor is the potential role of stellar rotation,
which could increase the brightness of the blue loop by as much as 0.7~mag
\citep{Milone17} and thereby imply an older age.
Note that the numbers of blue loop stars are expected to increase strongly with age,
as shown from the simulations in Figure~\ref{fig:agehist},
and are $\sim$~50\% more than observed for SC1 -- 
which may reflect uncertainties in the models from treatments of convective overshooting,
unresolved binaries and stellar rotation \citep{Barmina02,Costa19}.

\subsection{Implications of integrated colours}\label{sec:intcol}

A further consistency check on the age and metallicity solution
is to compare the predicted global colour of the
simulated cluster with the observed colour of SC1.
These are $\mathrm{(F475W-F606W})_0 = 0.258 \pm 0.014$
(where the uncertainty represents the stochasticity from 25 different simulations)
and $0.22 \pm 0.01$, respectively, 
which are very close\footnote{The CTIO and HSC colours are effectively slightly redder because of the presence of the embedded star St1.  This led to an initial impression of a negative age gradient between the inner and outer parts of the cluster (using resolved stars from HSC), but the {\it HST} data now demonstrate that there is no significant gradient.}.
The predicted colour for a younger age and higher metallicity (250 Myr; $Z=0.008$)
is also almost the same 
-- so here it is the CMD that provides the high-precision stellar population measurement.
We could in principle obtain stronger constraints from the integrated light measurements by using the wider wavelength range provided by the shallower {\it HST} photometry discussed in Section~\ref{sec:ILphot} (see Table~\ref{tab:clust_props}).
Comparing to predicted colours from PARSEC for our CMD-based solution,
the observed flux in the ultraviolet (F225W and F275W) is much too low, by $\sim$~0.3 mag.
Reasonable agreement could be obtained with older ages (e.g.\ $\sim$~500~Myr) 
and/or higher metallicities (e.g. $Z \sim 0.008$), or possibly by allowing for variations
in the Galactic extinction model (e.g.\ with $R_V < 3.1$).
However, all of these solutions would increase the F814W flux beyond what is observed.
Furthermore, we have checked predicted colours from Flexible Stellar Population Synthesis
\citep{Conroy09} and these are dramatically different from the PARSEC predictions (extremely bright in the ultraviolet).  Therefore we consider further analysis of the integrated-light colours to be beyond the scope of this paper.

\subsection{Age spread}\label{sec:dage}

Besides the mean magnitude of the blue loop marking the mean age of the stars,
the magnitude {\it width} can be translated to an age spread for the stars.
The strength of this constraint is unique to the CMD approach
(as compared to spectroscopy) and provided the main motivation for these
{\it HST} observations.
As shown by Figure~\ref{fig:agehist}, the standard deviations of the observed and 
simulated peaks are similar, at $\sigma_V = 0.20 \pm 0.02$ mag and $0.18 \pm 0.03$ mag, respectively. 
Formally, this means no more than $\sigma_V \simeq 0.15$~mag spread in the underlying ``isochrones''
(allowing for the uncertainties in $\sigma_V$),
which translates
to an upper limit of $\sim$~50~Myr on the total age spread, or equivalently
on the delay for any substantial second burst of star formation.
This constraint is relatively insensitive to the systematic uncertainties discussed above in the mean age determination, and in fact
also provides evidence against a large spread in stellar rotation
(unless there are subpopulations with both age and rotation differences that conspire to cancel out in the blue loop magnitudes).

There is no indication of a much older population ($> 1$~Gyr)
that might be underlying a ``frosting'' of younger stars.
These old stars would be seen on the upper red giant branch (RGB) at $V \sim$~25--27
and $g-V \sim 1.0$.  
Such stars are abundant in the outer disc or halo of NGC~247,
as we find in CMDs covering larger areas of the {\it HST} image
(as will be shown in Section~\ref{sec:filament}).
For a quantitative comparison,
there are $\sim$~300 RGB stars in the magnitude range $V = $~25--27 
in the 1600 arcsec$^2$ region surrounding SC1,
so we expect $\sim$~9 of these stars to appear associated with SC1 by chance.
There are 13 stars observed,  
which is fully consistent with a projection effect plus 
measurement scatter from the young stars (see right panel).  
Of course, other features (young or old)
could in principle be hiding inside the central 7~pc of SC1 that are too crowded for a CMD.

\subsection{Bright red straggler stars}\label{sec:strag}

We do not detect any candidate asymptotic giant branch (AGB) stars 
(brighter and redder than the blue loop)
in the $\sim$~100--300~Myr range, although the simulation shows that these would be
relatively rare.
On the other hand, the observations contain three bright stars 
in the outer annulus (and several more in the inner annulus)
that are not predicted by our preferred model solution
(perhaps one or two in the inner annulus could be from photometric scatter).
These stars are in the range of $V =$~22.7--24.5 and $g-V =$~0.2--1.0, 
and are marked in Figure~\ref{fig:syntimg} as small red circles.
These are generally too bright to be attributed to measurement errors or blends, according to our artificial star tests,
and almost all of them have good ``sharp'' values as well as good $\chi$ values.
Similar objects are also very rare in the surrounding areas, and we expect only
$\sim$~0.1 to appear projected on SC1 by chance. 
We can therefore comfortably assume that these ``red straggler'' stars are associated with SC1.

In principle, the red stragglers could be blue loop stars 
from younger sub-populations ($\sim$70 and $\sim$150~Myr),  
but then we do not observe the corresponding blue stragglers
that would be a bright extension of
the main sequence.
An alternative explanation
is the effects of binaries that are not included in our simulations -- 
either by appearing as unresolved,
bright single stars, or by physically merging into more massive, luminous stars
(see e.g. Section~3.5 of L11).
The latter is more realistic for providing the required brightness boost,
although merger products may also be more likely to become blue rather than red stars.

A final possibility is stars leaving the AGB phase on their way to becoming planetary nebulae, the latter of which have been found in young massive clusters (YMCs; \citealt{Larsen06}).
Here the problem is the extremely rapid (tens of years) transition timescale expected (e.g. \citealt{Miller16}), 
so that observing even one star in this phase would be unlikely, much less $\sim$~6 stars.
In summary, none of the explanations for the red stragglers seems satisfactory, 
which underlines the utility of young star clusters for providing constraints
on poorly understood aspects of stellar evolution.

Not included in Figure~\ref{fig:CMD} is the even
brighter, redder star St1 with $(g-V)_0 = 1.46, V_0 \sim 21.42$,
since our measured redshift (Section~\ref{sec:spec}) indicates 
that it may not be directly associated with SC1.
The velocity difference is 
$18 \pm 7$~km~s$^{-1}$
while the expected escape velocity of SC1 is $\sim$5~km~s$^{-1}$.
Even so, it seems a rare coincidence that this star is so close in both position
and velocity to SC1, and there may be alternative explanations.
Perhaps the star was until recently a member of SC1,
and is now escaping -- either as a component of the surrounding tidal debris
(Section~\ref{sec:filament})
or through high-velocity ejection \citep{Lennon18} --
but has not yet left the scene.
In this case,
the high luminosity of this star would be even harder to explain
than the other red stragglers in the context of SC1.
This star was also bright enough to detect in the shallow F814W imaging
(called $I$-band for short),
which provides an opportunity for additional diagnostics using a 
$(g-V)$ versus $V-I$ colour--colour diagram.
We estimate 
$(V-I)_0 \sim$~2.3--2.7 using aperture photometry.
Comparing PARSEC predictions for giant-branch stars with a range of ages
(to allow for stellar mergers), the observed $V-I$ colour is too red.
On the other hand, a foreground main-sequence dwarf star with a metallicity
of $Z \sim 0.008$ does fit the colours, suggesting the similar velocity to SC1 is
simply a coincidence.

\subsection{Cluster mass}\label{sec:mass}

To estimate the stellar mass of SC1, we try two complementary approaches.
The first is to self-consistently begin with the star counts from the simulated model that matches the observations (total magnitude and CMD morphology).
We sum up the initial stellar masses that are the inputs to this model, up to the main sequence turn-off mass of $3.1 \mathrm{M}_\odot$, using a Kroupa IMF.
We assume that all of the higher mass stars have expelled most of their mass from the cluster and left stellar remnants,
with a mean mass fraction of $\sim$~15\%.
This leads to a model mass-to-light ratio of $M/L_V \simeq 0.21$ in Solar units, and
to a final stellar mass of $M_\star \simeq (8.9\pm0.5)\times10^4 \mathrm{M}_\odot$.
Here the uncertainty comes from the photometric normalization and from the distance;
the uncertainty in $M/L$ from the stellar population modelling is difficult to
estimate but may be comparable in magnitude.

The second approach is to use a more general stellar population synthesis model 
\citep{Bruzual03} 
to calculate the passive fading expected between the
current age and $\sim$~10~Gyr in the future
(assuming no stars escape the system).
For an age of 316~Myr and $Z=0.004$, this is 2.8 mag in the $V$-band, 
leading to a future absolute magnitude of $M_V = -6.5$.
If we then assume a conventional $M/L_V \sim 2$ for old
star clusters, this leads to a mass of $M_\star \sim 7\times10^4 \mathrm{M}_\odot$, which is 
close to the value from the first approach.
The expected velocity dispersion of SC1 given its size and mass is
$\sigma \simeq 2$~km~s$^{-1}$.

\section{Discussion}\label{sec:disc}

Here we discuss various implications of the results derived in Section~\ref{sec:photres}.
Section~\ref{sec:class} considers how SC1 relates to other star clusters in size--mass splace.
Section~\ref{sec:mp} connects with multiple populations and age spreads in other clusters.
Section~\ref{sec:context} explores the formation history of SC1 in the context of its host galaxy, along with a complex of concurrently formed material (a stellar filament and lower-mass extended clusters).
Section~\ref{sec:form} summarizes general implications for the formation of low-density clusters.

\subsection{Classification}\label{sec:class}

\begin{figure*}
	\includegraphics[width=15cm]{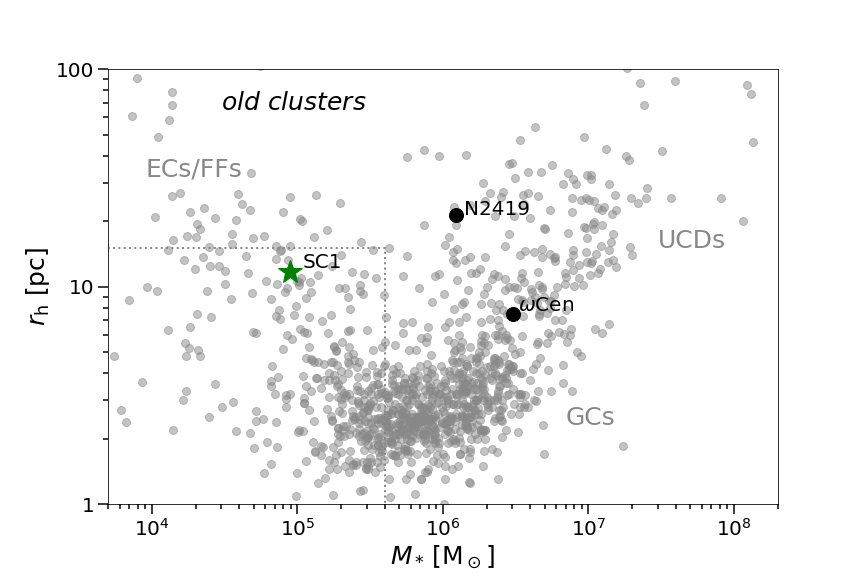} 
	\includegraphics[width=15cm]{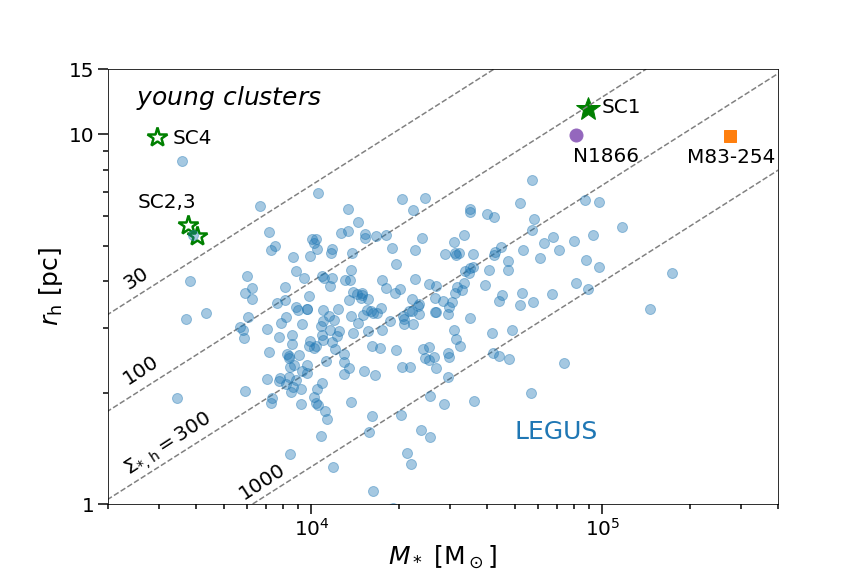} 
    \caption{Compendium plots of size (projected half-light radius) versus stellar mass
    for star clusters.
    {\it Top:} Old clusters,
    including globular clusters (GCs), ultracompact dwarfs (UCDs)
    and extended clusters (ECs) / faint fuzzies (FFs).
    NGC~247-SC1 is marked, along with
    several noteworthy MW objects ($\omega$~Cen, NGC~2419, M54).
    SC1 is intermediate in size and mass to compact, classical GCs and the
    most diffuse ECs and FFs.
    Note that star clusters and ultra-faint dwarf galaxies (UFDs) overlap in the upper left region of this plot
    and are difficult to disentangle, so the data shown likely include some contamination from UFDs.
    {\it Bottom:}
    Young clusters with ages of $\sim$~300~Myr 
    (note this panel is zoomed in relative to the top one, as indicated by the
    dotted lines). 
    Diagonal dashed lines show constant values of effective surface density, as labelled, in units of $\mathrm{M}_\odot$~pc$^{-2}$.
    Small blue circles show data from the LEGUS survey of nearby galaxies, as compiled by \citet{Brown21}.
    SC1 is marked by a large filled green star symbol:
    it is an outlier from almost all other young clusters in the plot.
    The three additional low-mass star clusters (SC2, SC3, SC4; see Section~\ref{sec:lowmass}) are marked by
    smaller open green star symbols;
    these also have relatively low densities.
    Other large symbols mark
     N5236-254 from the galaxy M83
     and NGC~1866 from the LMC.
    }
    \label{fig:uber}
\end{figure*}

Figure~\ref{fig:uber} places the size and stellar mass of SC1 in context with {\it old} star clusters observed in the nearby Universe, based on the compilation of \citet{Brodie11}, with subsequent updates\footnote{\url{https://sages.ucolick.org/spectral_database.html}.
A key asset of this catalog is that the objects have confirmed distances.}.
These literature data have $V$-band magnitudes that we convert to stellar mass
with an approximate mass-to-light ratio of $M/L_V = 2$.
Here it is important to recognize that there are strong selection effects in the sample, so the relative frequencies of objects in different parts of the diagram should be viewed with care, but it does provide a useful view of the range of parameter space that is known to be occupied by clusters.

SC1 resides in a well-populated region of size--mass space,
in between classical GCs and the most diffuse clusters such as the
Palomar clusters in the MW halo and most of the ECs in the M31 halo \citep{Huxor14}.
Other old stellar systems with similar properties include
FFs in the S0 galaxy NGC~1023 \citep{Larsen00b,Brodie02},
the MW halo clusters NGC 5053 and NGC 5466,
the Large Magellanic Cloud (LMC) cluster NGC~2257,
the Small Magellanic Cloud cluster NGC~339, 
cluster C2 in the Local Group dwarf NGC~6822 \citep{Hwang11},
halo cluster B in M33 \citep{Cockcroft11}
and a handful of halo clusters in M31 such as H15 and PAndAS-14.
Additional, similar objects with less secure ages and distances include FFs
in the interacting S0 galaxy M51B (NGC~5195; \citealt{Lee05,Hwang08})
and diffuse star clusters in Virgo and Fornax cluster early-type galaxies
\citep{Peng06,Liu16}.
In contrast, nuclear star clusters observed in the $\sim 10^5 \mathrm{M}_\odot$
mass range have much more compact sizes, with $r_{\rm h} \sim$~1--5~pc \citep{Pechetti20}. 

There is not an established distinction between ECs and FFs since they were discovered and named in different contexts, and continue to be discussed independently in the literature.
It does seem likely that there are two or more classes of diffuse star cluster with fundamentally different origins that happen to overlap in size--mass space.
A working definition that we adopt here is that ECs are metal-poor clusters found in dwarf galaxies and in the haloes of giant galaxies,
while FFs are metal-rich clusters associated with the discs of giant galaxies.
With [Fe/H]~$\sim -1$ as a very rough metallicity boundary, we then classify
SC1 as an FF, which is further supported by its tentative association with the
disc of its host galaxy.
We note that all of the ECs/FFs with CMDs observed previously have turned out to be
old and very metal-poor (e.g.\ \citealt{Huxor05,DaCosta09}),
and hence can be classified as ECs -- making SC1 the first FF with its CMD studied.

Next, we compare SC1 to star clusters of similar age
in the bottom panel of Figure~\ref{fig:uber}.
The primary data source is the LEGUS survey of 31 nearby star-forming galaxies, 
with cluster properties derived by \cite{Brown21}, constituting
the largest, unbiased high-quality sample of young cluster data available.
We select clusters with estimated ages of 300~Myr
and with ``reliable'' masses and radii ($\eta > 1.3$ in our equation~\ref{eq:EFF}),
and omit those from the galaxy NGC~1566 owing to the highly uncertain distance.
The typical sizes range from $\sim$~2 to 5~pc at masses of $\sim 10^4 \mathrm{M}_\odot$,
to $\sim$~3 to 7~pc at $\sim 10^5 \mathrm{M}_\odot$,
with median stellar surface densities of
$\simeq 300$~$\mathrm{M}_\odot$~pc$^{-2}$.
The large size of SC1 (and equivalently low density of $103\pm 10$~$\mathrm{M}_\odot$~pc$^{-2}$)
makes it an outlier in this context.

A well-studied galaxy missing from the LEGUS sample is the giant spiral M83.
One of the more extensive studies of its young clusters found that for
ages of 100--200~Myr and masses of $\sim4\times10^4 \mathrm{M}_\odot$,
the sizes are $\sim$~2--8 pc \citep{Ryon15}, comparable to the LEGUS results.
However, M83 also hosts a cluster that is somewhat analogous to SC1:
N5236-254, with an age of $\simeq$~280 Myr,
$M_\star \simeq 3\times10^5 \mathrm{M}_\odot$ and $r_{\rm h} \simeq 10$~pc
\citep{Larsen04,Larsen06}.
Intriguingly, it also lies towards the outskirts of its host galaxy,
at a galactocentric radius of $\sim$~7~kpc near the end of a spiral arm. 
Even so, it has a $\sim$~5 times higher density than SC1, similar to  
the lower mass clusters.

The LMC is known to host extended stellar clusters with a range of ages,
although their sizes can be difficult to define owing to model fits that
do not converge at large radii.
Considering the sample from \citet{McLaughlin05}, there is none with
a comparable age to SC1 and a well-constrained large size (e.g.\ $\eta > 1.3$).
Two that do have similar sizes and masses but different ages are
NGC~2121 (3 Gyr) and NGC 1866 (100~Myr, but see next Section 
with a more recent, older age estimate). 
The latter has a similar metallicity to SC1 and is perhaps the closest analogue;
interestingly, it also may be near the tip of a stellar ``arm'' of the LMC.

In summary, there are
no other obvious candidates for recently formed ECs or FFs
other than SC1 and perhaps NGC~1866.
A possible exception is the collection of
blue faint fuzzies around the S0 galaxy
NGC~1023 \citep{Forbes14}.  These may be young clusters, or they may be old and
metal-poor: more detailed observations would be required for confirmation.

The properties of the star cluster system of NGC~247 itself are also relevant
for putting SC1 in context.
The galaxy is relatively sparse in YMCs, 
which appears to be a natural reflection of
its low star formation rate  \citep{Larsen00a}.
Sizes were estimated for a few of these from ground-based data \citep{Larsen99}
but will need revisiting with higher-resolution imaging.
The old GC system remains relatively under-studied, with
\citet{Olsen04} having confirmed three clusters spectroscopically,
although there may be as many as $\sim$~60 GCs total.
These three clusters are all consistent with fairly high metallicities ([Fe/H] $\sim -1.0$ to $-0.6$),
old ages and sub-solar $\alpha$-element abundances.
They are projected at or just beyond the edge of the galactic disk, and appear to 
co-rotate with the disc.
Analysis of the HSC imaging indicates that these three are compact rather than extended clusters
like SC1
\citep{Santhanakrishnan16}.

\subsection{Multiple stellar populations}\label{sec:mp}

Multiple populations are a long-standing mystery in star clusters.
While old GCs are well approximated by coeval stars of a single age and metallicity,
they also show pervasive peculiarities in the star-to-star variations in light elements.
Most proposed explanations involve a second generation of stars that form from material
enriched by the first generation, on timescales of anywhere from $\sim$1 to $\sim$50 Myr.
Unfortunately, none of the explanations to date fits comfortably with the many
constraints provided by observations (see review by \citealt{Bastian18}).

Multiple populations have been generally detected in clusters down to ages of $\sim$~2~Gyr,
but not younger \citep{Cassisi20}
-- which is a major puzzle since there is no reason for the underlying physics or initial conditions to change for that age.
The suspicion has been that this incongruity reflects an observational limitation, e.g. by studying stars on the RGB rather than the main sequence, which seems to be confirmed by recent observations
\citep{Cadelano22}.
If YMCs do host ubiquitous multiple populations, they then serve as important test-beds
for the presence of age spreads which could be closely linked to abundance spreads.

In this vein, younger clusters do show widespread peculiarities in their CMDs,
such as extended main-sequence turn-offs
(e.g.\ \citealt{Mackey08,Milone09}; L11).
These were initially interpreted as age spreads of up to a few hundred Myr, but now appear more likely driven by other processes such as stellar rotation and mergers
(e.g.\ \citealt{Bastian09,Niederhofer15,Milone18,Kamann20,Wang22}).
Even so, rotation variations may connect to abundance variations in young clusters \citep{Pancino18}.
Furthermore, there are still signs of age spreads in some cases
\citep{Goudfrooij17,Costa19,Gossage19}, which motivates a wider inventory of CMDs in YMCs.

\begin{figure*}
	\includegraphics[width=17cm]{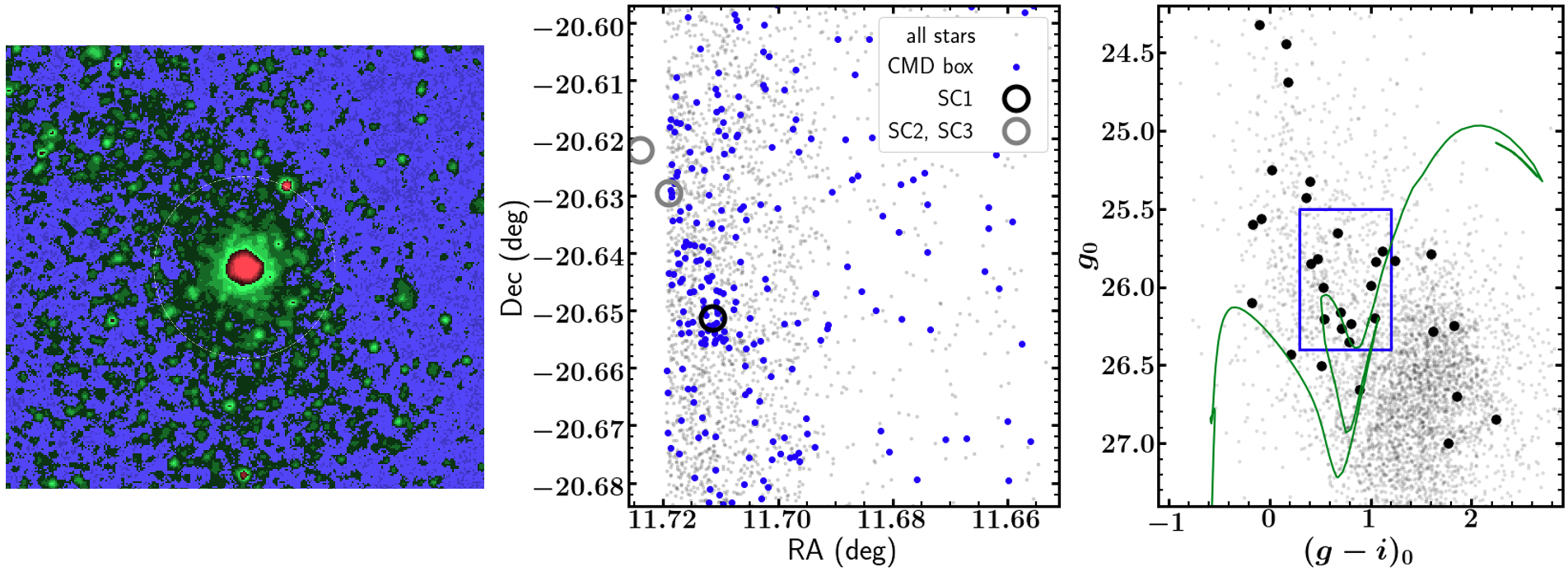}
    \caption{Filamentary feature extending through SC1,
    as seen in the HSC data.
    North is Up, and East is left.
    {\it Left:} 
    $g$-band image displayed with false colour for the sake of contrast, 
    where the dotted circle marks the expected $\simeq 9\arcsec$ tidal (150 pc) radius.
    There appear to be broad linear features extending outwards from just inside
    the tidal radius.
    {\it Middle:} 
    Spatial positions of star counts, 
    where small grey dots show all stars and
    larger blue circles show ``blue loop'' stars.
    The position of SC1 is marked by an open black circle, from which
    a linear feature in the star counts is seen to extend $\sim0.03^\circ$
    ($\sim$2~kpc) towards the Northeast.
    The {\it HST} image barely captures the filament but does not extend
    far enough to the West and South to pick up the density contrast with the
    surrounding regions (see Figure~\ref{fig:overview}).
    Two low-mass clusters (SC2 and SC3) are marked by open grey circles.
    The disc of NGC~247 begins at the left side of this plot,
    where crowding makes individual star photometry challenging
    (hence the ``empty'' region).
    {\it Right:}
    CMD of a $\sim$~50~arcmin$^2$ region in the halo of NGC~247 (grey dots). 
    The black points show stars selected within 10\arcsec\ of SC1.
    The green isochrone curve shown is for an age of 300~Myr
    and metallicity of $Z=0.004$.
    The blue rectangle shows the selection box used for blue-loop stars
    that are plotted in the middle panel.
    }
    \label{fig:stream}
\end{figure*}

SC1 represents a significant addition to the literature on CMD variations
-- as a young, high-mass cluster with a low density and a different formation environment from clusters studied in the MW and its satellites.

Our $\sim$~50~Myr upper limit on its age spread is in tension with the extended star formation histories required in AGB-polluter scenarios \citep{DErcole10}, although the low escape velocity of SC1 already
makes it unlikely to retain stellar ejecta.
As discussed in Section~\ref{sec:dage}, we also can exclude
significant spreads in stellar rotation.
Whether or not this cluster is completely free of multiple populations could be tested further with integrated light spectroscopic analysis of its abundances such as sodium.

The LMC cluster NGC~1866 
provides an intriguing comparison to SC1,
since its age, mass and size make it
a potential analogue to SC1 
(see Section~\ref{sec:class}).
This well-studied benchmark young cluster has a split main sequence that may reflect a combination of rotation and age variations \citep{Milone17,Dupree17}.
\citet{Costa19} analyzed a sample of Cepheids in this cluster and concluded that 
the stellar population overall most likely consists of a 
dominant fast rotating population with an age of $\simeq$~290~Myr,
and a secondary, relatively slow rotating population with an age of $\simeq$~180~Myr.
(along with small differences in the metallicity of [Fe/H]~$\sim -0.4$).
The fast rotating stars may be understood as inheriting angular momentum from their
initial gas cloud, but it is not understood how the cluster would re-ignite a second population.
Here we present a new puzzle of why the similar cluster SC1
does not likewise present evidence for multiple generations of stars.

\subsection{Host galaxy context: filament and low-mass clusters}\label{sec:context}

\subsubsection{SC1 orbit and stellar filament}\label{sec:filament}

Given the proximity of SC1 in projection to the outer rim of the disc of NGC~247
(Figure~\ref{fig:overview}),
we would like to know if there is a current or past physical connection.
Velocity is one potential clue, to test if an object is corotating with the disc
or is on a more random, halo-like orbit.
We use publicly available data from the VLA-ANGST survey \citep{Ott12} to map
out HI emission around NGC~247.
The cold gas disc of the galaxy extends out almost
to the position of SC1, with a projected separation of $\sim 20\arcsec \sim 350$~pc.
The HI velocity range in this area is $\simeq$~80--100~km~s$^{-1}$,
and the H$\alpha$ emission velocities appear similar
\citep{Hlavacek11}, 
in both cases close to
the SC1 velocity of 
$112 \pm 5$~km~s$^{-1}$\footnote{As discussed in Section~\ref{sec:spec}, this velocity uncertainty may be underestimated, and indeed preliminary analysis of a lower signal-to-noise but higher resolution spectrum from Keck/HIRES suggests a value of $\simeq$~95~km~s$^{-1}$.}.
An association between cluster and disc appears possible
(similar to the apparent co-rotation of the three known GCs),
although SC1 may also
be on a more random orbit whose line-of-sight velocity just happens to 
be similar to the disc's
(we revisit gas associations below).
If we assume SC1 is in the disc plane with an inclination of 76$^\circ$, 
we can deproject its position to give its true galactocentric distance as $r_{\rm G} \simeq$~14~kpc.

This inferred distance allows us to calculate the expected tidal radius $r_{\rm t}$ for SC1:
\begin{equation}
    r_{\rm t} = \left(\frac{GM r_{\rm G}^2}{2v^2}\right)^{1/3} ,
\end{equation}
where $M$ is the cluster mass
and $v$ is the circular velocity of the host
(see \citealt{Baumgardt10}).
With $v\simeq$~110~km~s$^{-1}$ \citep{Lelli16,Ponomareva16},
we find $r_{\rm t} \simeq 145$~pc~$\simeq 8\farcs5$. 
If SC1 is not actually in the disc plane, and we allow for a wider range
of plausible distances $r_{\rm G} \sim$~10--25~kpc, this 
leads to $r_{\rm t} \sim $~120--220~pc~$\sim$~7\arcsec--13\arcsec.

Given the extended nature of SC1 and its relative proximity to its host galaxy, 
we next test for indications of extra-tidal stars or other non-equilibrium features.
Visual inspection of the HSC $g$-band imaging suggests the cluster stays relatively
round and asymmetric out to $\sim$~5--6\arcsec~$\sim$~100~pc,
but beyond this radius shows signs of broad, extended features to the Northeast and Southwest
(see left panel of Figure~\ref{fig:stream}).
This transition is suggestive of a slightly smaller tidal radius than our default value,
which may in turn imply that SC1 is at a distance of $r_{\rm G} \sim$~8~kpc
and is {\it not} associated with the disc.
This smaller distance would imply the cluster is well out of the disc plane into the inner halo,
at a height of $\sim 3$~kpc. 

\begin{figure*}
	\includegraphics[width=19cm]{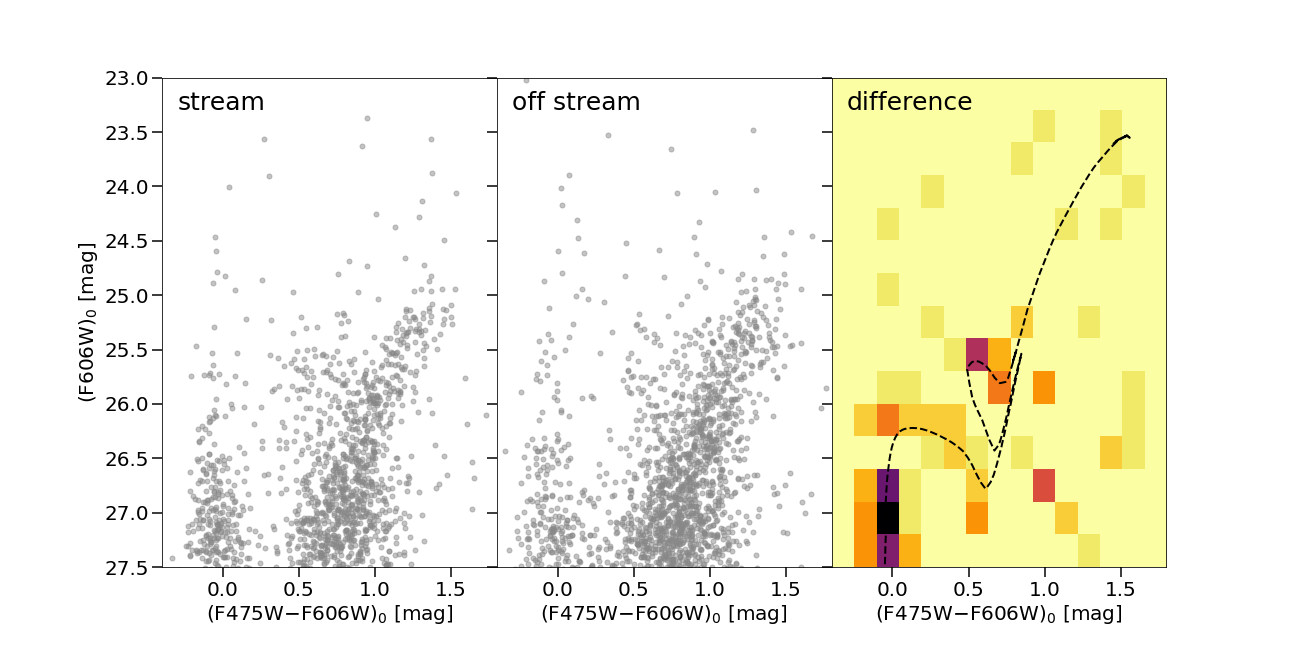}
    \caption{CMDs of point sources from 0.5~arcmin$^2$ regions of
    the {\it HST} image.
    {\it Left}: Filament associated with SC1.
    {\it Middle:} Off-filament area to the East.
    {\it Right:} On/off filament difference, expressed as a Hess diagram,
    where darker shading represents higher densities.
    The dashed curve shows the $Z=0.004$, 316~Myr isochrone that matches the SC1 data
    (Figure~\ref{fig:CMD}).
    The filament contains an excess of both blue-loop and main-sequence stars,
    whose locations in the CMD are consistent with SC1.
    }
    \label{fig:cmd3}
\end{figure*}

The Southwest feature appears to terminate fairly sharply at a distance of $\sim 25\arcsec$ ($\sim$~400~pc),
while the Northeast feature extends a much longer distance of $\sim 2\arcmin \sim$~2~kpc,
with a width of roughly $\sim 15 \arcsec \sim 250$~pc, until it overlaps with the disc.
Thus it appears that SC1 is located around the end of a stellar filament
that could be a tidal stream or a very faint spiral arm.

For more precise mapping out of the filament, we turn to star counts from HSC (Section~\ref{sec:hsc}),
using the stars associated with SC1 (within a radius of 10\arcsec)
to motivate a colour--magnitude selection area for the filament
(see right panel of Figure~\ref{fig:stream}).
This procedure assumes that the stellar populations in the filament and the cluster
are similar, which will be tested further below.
We use colour and magnitude ranges of 
$(g-i)_0=$~0.3--1.2 and $g_0 =$~25.5--26.4,
respectively, which correspond approximately to the empirical blue-loop region.
The spatial distribution of these ``blue-loop'' stars is shown in the
middle panel of Figure~\ref{fig:stream},
which shows a narrow, linear feature extending to the Northeast of SC1
and confirms the visual impression from the original image.
If the selection box is moved to bluer or redder colours (e.g.\ capturing RGB stars),
then the density map no longer shows the filamentary feature.
The stellar population of this feature will be examined below in more detail using {\it HST},
while here we test the statistical significance of the Southern truncation
using HSC with its larger field of view.  
We place 740 arcsec$^2$ rectangular apertures along the filament,
finding that the number of blue-loop stars remains constant at 15--16
until the truncation is reached at $\sim$~25--30 arcsec from the center of SC1.  Beyond this point,
the number in the aperture is 1--3 stars, i.e. a decrease at the 3$\sigma$ level.

We next check multi-wavelength archival imaging from NED for any other signs
of the filament.  
It does appear to be detected but ill-defined in CTIO $B$-band imaging from
the {\it Spitzer} Local Volume Legacy survey \citep{Cook14a}.
Similarly, it is marginally visible in luminance-filter imaging \citep{Rich19}.
In {\it GALEX},
there appears to be slight excess emission roughly in the same area as the filament, 
both in FUV and NUV,
but it may not be well aligned with the HSC feature
(in contrast, SC1 itself is a strong ultraviolet source).

We also revisit the VLA data to search for lower-density gas 
around SC1 and the filament.
We do indeed detect signs of widespread gas not reported in previous HI observations, but at faint levels that are difficult to map out
over large spatial scales given the minimum VLA baseline, and
there is no clear detection of a linear feature associated with the stellar filament.
We find typical gas mass densities of $M_{\rm HI} \sim 2\times10^6 {\rm M}_\odot$~kpc$^{-2}$,
at velocities of $\simeq$50--100~km~s$^{-1}$.
This fairly uniform gas distribution at disc-like velocities most likely represents an extension
of the disc, with any remaining cold gas associated with the filament either present at a lower mass level or dispersed to other parts of the galaxy.
Star formation rates at these low densities in outer discs are expected to be extremely low,
$\Sigma_\mathrm{SFR} \sim 2\times10^4 \mathrm{M}_\odot$~Gyr$^{-1}$
\citep{Bigiel10}, so these gas detections should not represent the formation sites for SC1 and the stellar filament --
although there is precedent for the apparent formation of massive star clusters in the low-density outskirts of M83 \citep{Dong08}.

\begin{figure*}
	\includegraphics[width=17.8cm]{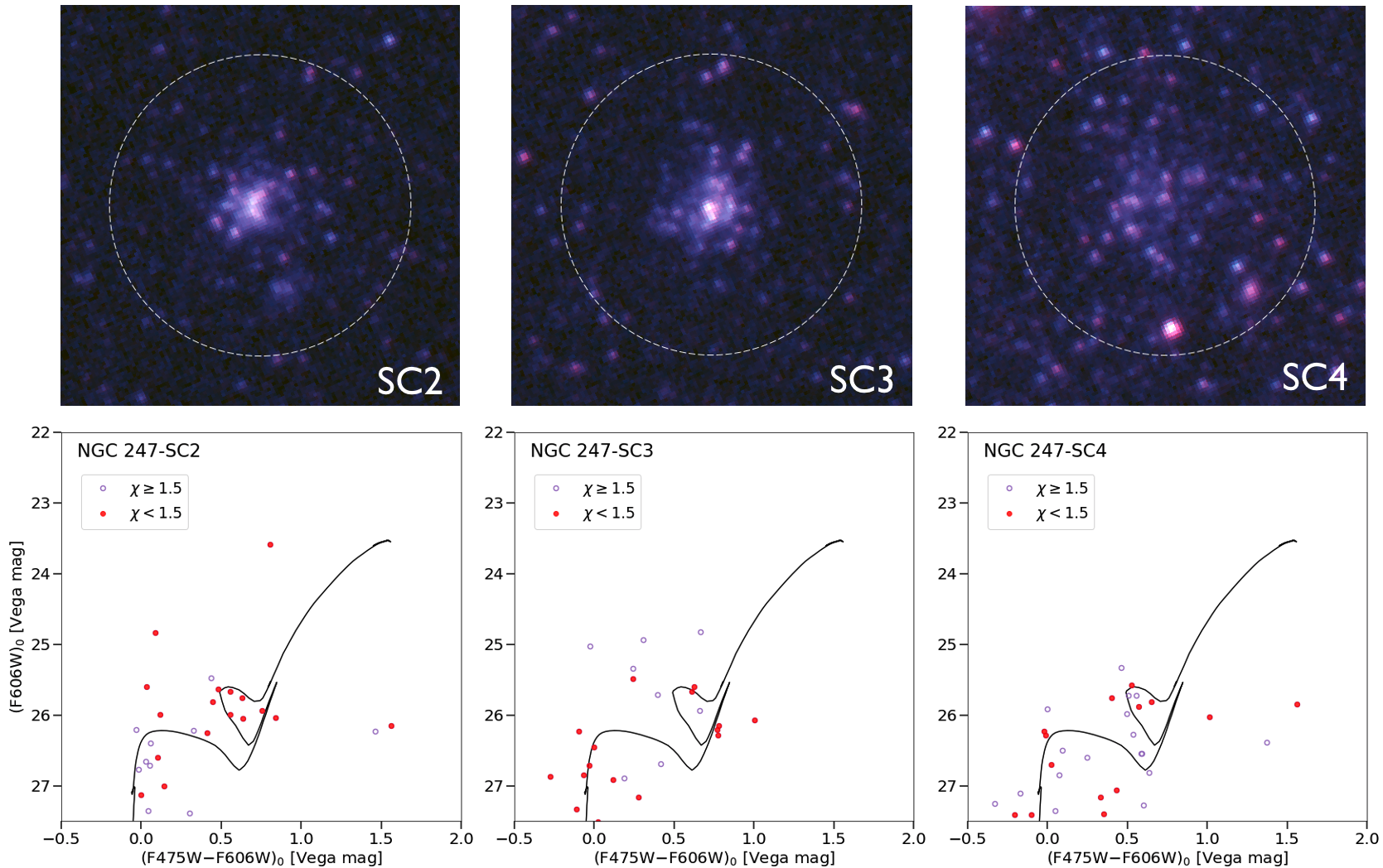}
    \caption{Low-mass clusters SC2, SC3 and SC4 (left to right)
    identified in {\it HST}/WFC3 imaging.
    {\it Top panels} show colour thumbnails using the F475W and F606W filters.
    North is up and East is left.
    Large dotted circles are drawn for reference, with a radius of $1\farcs8 \simeq 31$~pc.
    SC2 and SC3 are in the filament extending northwards from SC1.
    SC4 is in an outer section of the host galaxy disc.
    {\it Bottom panels} show the CMDs for the clusters within a radius of
    $0\farcs8 \simeq$~14 pc,
    using the same axis ranges as in the SC1 CMD in Figure~\ref{fig:CMD}.
    The higher and lower quality measurements are indicated by red filled and
    open purple circles, respectively.
    The curves show a 316 Myr, $Z=0.004$ isochrone, which appears
    consistent with the data for all three clusters.
    Nearby control fields of similar areas would show only $\sim$~3 stars in the CMD in the cases of SC2 and SC3, 
    and $\sim$~12 stars in the case of SC4 (most of these away from the isochrone).
    }
    \label{fig:thumb}
\end{figure*}

The {\it HST} imaging is less effective for capturing the density 
contrast between the filament and the surroundings, 
owing to the smaller field of view,
although the filament is detectable in
carefully smoothed maps of young star counts, 
particularly close to SC1. 
Instead, {\it HST} is most useful for examining the stellar population of
the filament in detail, as shown by Figure~\ref{fig:cmd3}.
Here we have selected equal areas of the {\it HST} image (0.5~arcmin$^2$),
both on (left panel) and off (middle panel) the filament (immediately to the East).
The differences between the panels are subtle, as both fields show a dominant RGB population as well as a main sequence extending to bright magnitudes.
The filament field does show a modest excess of stars in the blue-loop
region at $(g-V)_0 \sim 0.7, V_0 \sim 25.7$.

To view the differences more clearly, we construct a Hess
CMD binned-density diagram of each of the two fields and then subtract them in order to generate
a difference map (right panel).
Here we display negative values as zero (light yellow)
in order to avoid being dominated by the radial gradient in RGB density.
Two primary features emerge that are insensitive to the binning scheme:
an excess of main sequence stars at $V \sim$~26--27.5, 
and an excess of blue-loop stars.
The locations of these features in the CMD are very similar to those
of SC1 -- as can be seen by using the isochrone curve as a reference, 
and as we have confirmed from examining their luminosity functions 
and comparing to Figure~\ref{fig:agehist}.
To test the statistical significance of this result, we carry out 25 bootstrap resampling iterations of the two fields, and find that in every case, the Hess diagram shows a clear overdensity along the SC1 isochrone.
These features also confirm the result from the HSC data (Figure~\ref{fig:stream})
that the filament consists of a population of young stars with no sign of
underlying old stars.
There are also hints of an excess of red straggler stars
along the filament and in the general vicinity of SC1,
but the numbers are too low to be sure.

\begin{table*}
	\centering
	\caption{Properties of low-mass star clusters around NGC~247 -- see Table~\ref{tab:clust_props} for details of columns.}
	\label{tab:smallclust}
	\begin{tabular}{lccccccccc} 
		\hline
		Name & RA & Decl. & (F606W)$_0$ & (F475W$-$F606W)$_0$ & $g_0$ & $(g-i)_0$ & $M_V$ & $M_\star$ & $r_{\rm h}$ \\
		 & (J2000) & (J2000) & (Vega mag) & (Vega mag) & (AB mag) & (AB mag) & (Vega mag) & ($\mathrm{M}_\odot$) & (pc) \\
		\hline
		NGC 247-SC2 & 11.7191 & $-$20.6295 & $21.86 \pm 0.05$ & $0.17 \pm 0.05$ & $21.85 \pm 0.13$ & $-0.14 \pm 0.25$ & $-5.82 \pm 0.08$ & $3800 \pm 200$ & $5.7 \pm 0.5$ \\
		NGC 247-SC3 & 11.7240 & $-$20.6221 & $21.77 \pm 0.05$ & $0.23 \pm 0.05$ & $21.66 \pm 0.20$ & $0.58 \pm 0.23$ & $-5.90 \pm 0.08$ & $4000 \pm 200$ & $5.3 \pm 0.5$ \\
		NGC 247-SC4 & 11.7348 & $-$20.6190 & $22.05 \pm 0.20$ & $0.29 \pm 0.20$ & $22.53 \pm 0.21$ & $0.31 \pm 0.29$ & $-5.55 \pm 0.21$ & $2900 \pm 600$ & $9.8 \pm 1.7$ \\
		\hline
	\end{tabular}
\end{table*}

To estimate the stellar mass of the filament, we use stars from the
upper main sequence and the blue loop as tracers, calibrating the
ratios between these stars and total stellar mass using the outer
annulus of SC1.
Over a filament area of $\simeq 0.7$~arcmin$^2 \simeq 0.7$~kpc$^2$, 
we use this approach (after correcting for background levels) to
find $M_\star \sim 10^4 \mathrm{M}_\odot$,
i.e.\ a factor of $\sim$~10 less massive than SC1.
The equivalent surface brightness is
$\mu_V \simeq$~29.4~mag~arcsec$^{-2}$, which is at or beyond the
limit of the very deepest studies of extragalactic streams
in integrated light (see summary in Section~3.3 of \citealt{Duc15}).
After 10 Gyr of fading, and if still intact, the filament would have
$\mu_V \simeq$~32~mag~arcsec$^{-2}$, and any such features in the haloes
of galaxies beyond the Local Group would be essentially undetectable.

\subsubsection{Low-mass star clusters}\label{sec:lowmass}

We next carry out an initial search for more star clusters 
that may be coeval with SC1, since star clusters normally form together in a cohort with a
characteristic mass distribution (e.g.\ \citealt{Larsen09}).
SC1 is near the characteristic upper-mass cutoff of this distribution, and
thus we expect many more lower-mass clusters to have formed --
although after 300~Myr, many of them may have been disrupted or else
diverged in their orbits to different parts of the galaxy.
We search for clusters in the Western half of the {\it HST} image (away from the galaxy disc),
both by eye and by using star counts,
using a CMD box to select blue-loop stars as done with the HSC data, and now with
box boundaries of $(g-V)_0 = $~0.35--0.85 and $V_0 = $~25.4--26.7~mag.
There are three obvious clusters of stars, all of them toward the NNE of SC1,
at distances of 1\farcm4--2\farcm4 = 1.4--2.4 kpc
(see thumbnail images and CMDs in Figure~\ref{fig:thumb}). 
Two of these appear to be embedded in the filament 
(see right panel of Figure~\ref{fig:stream}), 
and one is in a protrusion of the NGC~247 disc near the ``top'' of the filament.
In the HSC imaging, they would be much more difficult to identify, as small
smudges with more stochastic global colours.

We dub the three low-mass clusters SC2, SC3 and SC4, in order of proximity to SC1,
and summarize their properties in Table~\ref{tab:smallclust}.
Here we have used aperture-photometry profiles to derive their {\it HST}-based
total magnitudes, colours and
sizes, as done for SC1 in Section~\ref{sec:ILphot}, but now with a maximum radius of 30~pixels (1\farcs2).
Their colours are all consistent with that of SC1, as expected for a coeval population.
We have also derived HSC $g,i$ photometry using the procedures described in Section~\ref{sec:ILphot}, although we had to use a small 1\farcs8 aperture for SC4
to avoid including flux from nearby bright stars.

To derive the cluster masses, we assume the same $M/L_V=0.21$ as for SC1.
We plot their sizes (5--10~pc) and masses (3000--4000~$\mathrm{M}_\odot$)
in the bottom panel of Figure~\ref{fig:uber},
where they appear as relatively low-density clusters (5--20~$\mathrm{M}_\odot$~pc$^{-2}$), 
at the upper size envelope of 
the LEGUS distribution, although this survey may be incomplete for such clusters. 
Analogous clusters in the Local Group with old ages are Pal~4, M33-D and PAndAS-45. 
We calculate expected tidal radii as done above for SC1, assuming $r_{\rm G} \sim 10$~kpc,
and find that they all have $r_{\rm t} \sim$~40~pc $\sim 2\farcs3$.
This means that the clusters should not be in imminent danger of disrupting,
and in the cases of SC2 and SC3 there are no obvious indications of extra-tidal material.
The very low density cluster SC4 appears to be particularly asymmetric, and its
boundaries are unclear since it is embedded in the disc.
This object seems the most likely to disrupt soon, 
although its location in the disc may also be only a projection effect.

The existence of these low mass clusters along the filament reinforces a picture
where all of this material including SC1 is connected,
and originated in the same star-forming event.
An alternative scenario where the filament represents tidal material lost from SC1 is undermined by the very strong asymmetry of the filament relative to the cluster
(i.e.\ missing either the leading or the trailing tidal tail, which would be very
hard to hide in projection).

\subsubsection{Connections to galaxy disturbances}

While this filament has not been previously reported, there are other signatures
of disturbances in the recent history of NGC~247.
A $\sim$~3~kpc ``void'' in the Northern half of the galaxy was examined in detail by
\citet{Wagner14} and suggested as the imprint of a nearly-dark subhalo impact.
\citet{Davidge21} subsequently mapped out the large-scale multi-wavelength properties of the
galaxy and determined that the void is an illusion created by an over-luminous spiral arm
on the Northern edge of disc, which in turn was most likely provoked by an external perturbation.
This would have occurred within the last few dynamical times of the disc,
which is $\sim$~100--300~Myr.
There are also large bubble-shaped regions of young stars in the disc with estimated expansion ages
of $\sim$150--250 Myr \citep{Davidge21}.
The galactic nucleus appears to have experienced a starburst $\sim$100--300 Myr ago
\citep{Kacharov18}.
All together, it appears there was a galaxy-wide disturbance $\sim$300~Myr ago
that continued until $\sim$100~Myr ago, followed by a period of relative quiescence
(NGC~247 is now $\sim 1 \sigma$ below the star-forming main sequence;
\citealt{Leroy19}).

The 300~Myr age of SC1 coincides with the beginning of the galaxy disturbance and suggests
a link between the two -- with SC1 being either the culprit or a by-product of
the event.
Its orbital period is likely in the range of $\sim$0.5--1~Gyr,
so it is plausible that SC1 was born in or near the galaxy disc 300 Myr ago,
before travelling to the present position.
The $\sim 10^5 \mathrm{\mathrm{M}_\odot}$ stellar mass in SC1 
is too low to induce the large-scale disturbances in the host galaxy,
considering the disc mass is $M_\star \sim 3\times 10^9 \mathrm{M}_\odot$. 
For SC1 to be responsible, there must have been a lot more accompanying material --
either in dark matter or in gas -- that is not observable now.

We thus consider first a conventional minor merger scenario involving a disrupting satellite galaxy, 
analogous to the Sagittarius dwarf interaction with the MW
(e.g.\ \citealt{Purcell11,Ruiz20}).
The lower mass of the host galaxy, NGC~247, would then imply a much fainter satellite galaxy perturber, given the steep stellar-to-halo mass relation for dwarf galaxies (e.g.\ \citealt{Wechsler18}).
A dark subhalo mass of $\sim 10^8 \mathrm{M}_\odot$ has previously been suggested for producing the void in NGC~247
(\citealt{Wagner14}; see also \citealt{Kannan12,Shah19}).
This would be the mass at impact, while the mass at infall before tidal
stripping of the dark matter halo would be
$\sim 10^9 \mathrm{M}_\odot$. 
Expectations for stellar masses in this regime are an area of active research and still
very uncertain, but current estimates are 
in the range of $\sim 10^3$--$10^5 \mathrm{M}_\odot$ \citep{Wheeler19,Applebaum21},
i.e.\ an ultrafaint dwarf (UFD).

A relevant observational example of a satellite system is NGC~2403: 
another nearby galaxy from the MADCASH survey with comparable mass and morphology to NGC~247.  
It has two known satellites, DDO~44 and MADCASH-1  \citep{Carlin19,Carlin21},
the former of which is disrupting,
with stellar masses of $2\times10^7 \mathrm{M}_\odot$ and $2\times10^5 \mathrm{M}_\odot$, respectively.
Either of these would have more than sufficient halo mass to produce the disturbances in NGC~247
(which incidentally raises the question of why NGC~2403 appears unscathed despite the recent pericentric passage of DDO~44). 
Another example is the irregular dwarf NGC~4449 whose starbursting activity was
unexplained until the discovery of a disrupting dwarf around it with a 
stellar mass of $2\times10^7 \mathrm{M}_\odot$ --
a so-called stealth merger \citep{Martinez12}.
These satellites are generally more massive than required for the NGC~247 disturbances,
but illustrate the idea that external perturbers can easily evade detection.

We are not aware of any satellite galaxies
around NGC~247 ---
other than two fairly massive ones that will be discussed below --
nor of any other tidal features in either gas or stars \citep{Westmeier17,Rich19}.
However, a UFD could have been missed -- almost certainly so if it were disrupted
(recall that SC1 would not be a direct remnant of the satellite,
owing to the single-burst recent star formation history and to its off-centre location in the filament.)
SC1 and the associated stellar filament and low-mass clusters might then have formed
either from disc gas from NGC~247 flung out after impact -- a ``galactic feather''
scenario \citep{Laporte19,Martinez21} -- 
or from cold gas belonging to the UFD.

There are indications that UFDs can retain gas in the field even if they quench very early
\citep{Janesh19,Applebaum21}, and perhaps they can also avoid ram-pressure stripping during infall to a low-mass galaxy.
If the precursor to SC1 harboured at least $\sim 10^6 \mathrm{M}_\odot$ of cold gas, this could
have been compressed during disc passage to form a massive cluster along with a train of
stars and lower-mass clusters.
An analogous case is the unique object Price-Whelan~1 (PW~1): a young ($\sim$120 Myr) open cluster
far out in the MW halo \citep{Price19}, which is thought to have formed from gas in the leading edge of the Magellanic stream around the time of disc crossing \citep{Bellazzini19}.
PW~1 is actually composed of at least three sub-clumps with masses of $M_\star \simeq$200--700~M$_\odot$
and sizes of $r_{\rm h} \simeq $~30--120~pc -- i.e.\ with even lower densities than SC1 and its associated clusters
(see again Figure~\ref{fig:uber}). 
Despite the similarities, though, there is a dramatic difference in outcomes between just
$\sim 10^3 \mathrm{M}_\odot$ of stars forming from $3\times10^7 \mathrm{M}_\odot$ of gas
\citep{Bruns05} in the case of PW~1,
and $\sim 10^5 \mathrm{M}_\odot$ in the case of SC1 and its likely less massive gas reservoir.

To distinguish between the above two possibilities -- star formation from gas in the disc versus gas in the satellite -- we may consider the [$Z$/H]$\simeq -0.6$ metallicity of SC1.
The disc metallicity for young stars ($\sim$15--40 Myr) was determined to be 
[$Z$/H]$\sim -0.3$ based on the colours of red supergiant stars
from ground-based imaging, and with little variation
with galactocentric radius \citep{Davidge06}.
The $\sim 0.3$~dex difference between these two measurements
may not be meaningful given the statistical and systematic uncertainties.
Estimating the disc metallicity from the area
covered by our {\it HST} pointing would also be a challenging exercise
beyond the scope of this paper.
For now, we consider the approximate match in metallicity to be indicative of a disc-related origin for SC1.
For example, a UFD origin would lead to a much lower metallicity
between [$Z$/H]~$\sim -3$ and $-2$ \citep{Simon19},
unless there was some mixing between UFD and disc gas (cf.\ \citealt{Tepper18}).

We consider next an alternative scenario, of a pure gas cloud with no dark matter that
interacted with the disc of NGC~247 and triggered its own internal burst of star formation
before going on to disrupt.
This HVC could either have an external origin or be re-accreted
after an outflow from NGC~247 
-- with the latter strongly preferred by metallicity considerations.
\citet{Davidge21} suggested HVCs as 
the triggers for the disk bubbles in NGC~247,
and \citet{Price19} suggested an HVC as an alternative origin for PW~1.
It is, however, unclear how plausible it is for
a low-mass galaxy like NGC~247 to host an HVC 
with a mass of $\sim 10^8 \mathrm{M}_\odot$ (as needed to explain the dynamical disturbances).

A third possible scenario is a fly-by (not an impact)
of a relatively massive satellite galaxy
that perturbed the disc of NGC~247 and triggered the formation of SC1 --
either in a galactic feather or via an HVC that was shed by the satellite and impacted the disc.
The provocateur galaxy would still be visible nearby:
assuming a mean relative velocity of $\sim$~100~km~s$^{-1}$,
after 300 Myr the distance travelled would be only $\sim$~30~kpc.
There are two known satellites that might fit this scenario,
with three-dimensional (3D) distances to NGC~247 that are constrained by a homogeneous
tip of the RGB method \citep{Dalcanton09}:
UGCA~15 (DDO~6) and ESO~540-G032, both with stellar masses of $\sim 3\times10^7 \mathrm{M}_\odot$ and classified as transition-type dwarfs that are gas rich
but with recently decreased star formation \citep{Weisz11}.
Their relative velocities and most likely 3D distances are fairly large
and suggestive of recent infall on wide orbits, but their minimum distances
are compatible with having a recent close approach to NGC~247 
-- particularly in the case of UGCA~15 (40 kpc).
Their stellar metallicities are in the range of [$Z$/H]~$\sim -2$ to $-1.5$
\citep{Sharina08,Lianou13}, which suggests they did not shed gas to form SC1 unless there was mixing with NGC~247 disc gas,
so the most plausible scenario is that
one of them excited a galactic feather in the disc of NGC~247,

A final scenario is that SC1 simply formed in the outer disc of NGC~247,
with a satellite interaction as in the previous scenario both triggering
the formation of SC1 through a density wave (cf.\ \citealt{Bush10})  
and clearing away the requisite gas (which might also have lower metallicity than the main disc).

In summary, we have examined several different scenarios
(minor merger, HVC infall, satellite fly-by, triggered in-situ formation)
that relate the origin of the SC1 complex
to other disturbances in NGC~247.
If the metallicities of SC1 and the host galaxy disc are indeed the same,
then probably the best explanation would be formation in a filament of disc
material that was dynamically perturbed.
Dynamical models of the interaction could provide further clarification,
and more work is also needed on confirming the metallicity of the NGC~247 disc.

\subsection{Formation mechanisms of low-density clusters}\label{sec:form}

Given the nature of SC1 as the first clearly identified young FF,
we consider briefly
how its formation history may connect to its size, 
and if there are any broader implications for the formation of low-density star clusters.
Star cluster formation through galaxy interactions and within tidal debris
has been well-studied (e.g.\ \citealt{Whitmore99,Boselli18,Fensch19}),
but any corresponding variations in cluster size have not.
The size trends for ordinary GCs are not fully understood, much less
those of clusters deviating from the average.
GCs are presumed to form from giant molecular clouds (GMCs), which
have surface densities of $\sim$100--1000~$\mathrm{M}_\odot$~pc$^{-2}$ 
depending on the galactic environment and the region within the cloud.
For a given environment and spatial scale, the GMCs have a near-constant
surface density, or equivalently a strong size--mass relation.
GCs, on the other hand, have near-constant sizes with surface densities
ranging from $\sim$100 to $\sim 10^4 \mathrm{M}_\odot$~pc$^{-2}$
(e.g.\ \citealt{Krumholz19}).
Young clusters appear to bridge this gap between GCs and GMCs, with a size--mass relation that corresponds to a relatively narrow range of surface densities
that is strongly peaked around 300~$\mathrm{M}_\odot$~pc$^{-2}$ 
(see lower panel of Figure~\ref{fig:uber})
and is a fairly universal trend
from galaxy to galaxy, despite the host galaxies having a wide range of star formation rates \citep{Brown21}.

While mechanisms that conspire to decouple GC sizes from their parent GMC sizes remain unclear, 
the origins of ECs and FFs with densities as low as $\sim 10$~$\mathrm{M}_\odot$~pc$^{-2}$ are even murkier.
One tidy explanation for ECs is that some star clusters started out compact but expanded to fill their
tidal radii, driven by internal dynamics
(e.g.\ \citealt{Gieles10,Madrid12}). 
However, not all clusters fill their tidal radii, and it appears that there is already a difference
at birth between compact and extended clusters \citep{Baumgardt10,Hurley10,Bianchini15}.

SC1 provides an important example of a massive cluster born large, as it has not lived enough Gyr for
the internal expansion processes to take effect.
If it is a typical progenitor of old FFs, then it may represent a
different formation pathway than for compact star clusters.
In addition, the conditions for this pathway would not be ubiquitous,
since there are strong differences between galaxies, even of the same type and environment,
in how many FFs they host \citep{Peng06}.
The FFs do generally appear associated with galactic discs, 
and could be considered as
massive open clusters, but this still begs the question: why do discs form some clusters
diffuse and some compact?  And why do some discs form {\it only} compact clusters?

One mechanism proposed for the formation of discy systems of FFs is
head-on galaxy mergers similar to the Cartwheel \citep{Burkert05}.
The resulting clumpy rings might be conducive to coalescence of small clusters into
large, low-density clusters, rather than forming the usual dense progenitors of GCs.
\citet{Elmegreen08} developed a related
model that explains FF formation through high Mach numbers in low density gas clouds,
such as may be found in collisional rings.
NGC~247 is not such an extreme example of a collision,
nor of a populous system of FFs (which host $\sim$20 objects; \citealt{Liu16}),
but the localized physical conditions leading to the formation of SC1
may have been similar to those.
More generally, SC1 provides fresh evidence that low-density cluster formation is somehow related to galaxy interactions -- a scenario that merits further study. 

\section{Conclusions}\label{sec:conc}

We have discovered and characterized a young faint fuzzy (FF) star cluster, SC1, around
the nearby low-mass spiral galaxy NGC~247.
This is a rare case of an FF found shortly after formation,
providing unique leverage on understanding the origins of low-density star clusters.
{\it HST} analysis of the CMD of SC1, assisted by parallel analysis of simulated cluster observations, indicates an age of $\simeq$~300~Myr and a metallicity
of [$Z$/H]$~\sim -0.6$, based on blue-loop stars with support from the main sequence turn-off,
and where the precision is limited by systematics.
The luminosities of the blue loop stars have no significant scatter relative to a simulated population from the isochrone, placing an upper limit on the age spread of $\sim$~50~Myr. 
SC1 thus differs from other young and intermediate-age clusters that have been claimed as having larger age spreads -- underlining the complexity of
piecing together the general puzzle of multiple populations.
There are several bright stars with intermediate colours that are associated with SC1 and that are difficult to explain with the single stellar population.
One possible explanation is that these are the product of post-main-sequence binary-star mergers.

The estimated stellar mass and size of SC1 are $M_\star = (8.9\pm0.5)\times10^4 \mathrm{M}_\odot$
and $r_{\rm h} = 11.7 \pm 0.5$~pc.
SC1 appears to be surrounded by local tidal debris and on larger scales
to be associated with structure of the same stellar population age:
a low-mass ($\sim 10^4 \mathrm{M}_\odot$) stellar filament
that is $\sim$~2~kpc long and that contains two lower-mass,
relatively large clusters
($M_\star \sim 4000 \mathrm{M}_\odot$, $r_{\rm h} \simeq$~5--6~pc), 
plus an even larger low-mass cluster ($r_{\rm h} \simeq$~10~pc) 
that may be embedded in the outer disc.

SC1 is close to the outer edge of the host galaxy disc in projection, and has a velocity that is similar to but significantly different from the disc velocity field.
Tidal radius arguments suggest that the cluster may be orbiting in the galaxy halo.
Whether currently located in the halo or the disc, the question remains of how such a relatively massive cluster of young stars ended up outside of the main star-forming regions of the disc.
We consider various explanatory scenarios for both SC1 and
its associated stellar filament and low-mass clusters, in conjunction with previous evidence for significant perturbation of the host galaxy.
Here a key constraint is that the metallicity of this new material is comparable to, and possibly slightly lower than, the young-disc metallicity.

Although the formation scenarios are still somewhat speculative, two general candidates emerge.
First is the origin of the SC1 complex in disc material that was either flung out of the disc by a galaxy interaction as a ``galactic feather,'' or was formed in a warped outer disc that subsequently lost much of its gas.
Here there may be some tension with the relative metallicities of SC1 and the disc.
The second scenario is the passage of
a gas-rich ultrafaint dwarf galaxy through the disc to produce a gas streamer that was a mixture of dwarf gas and disc gas.
This streamer experienced enough shocking to form stars and clusters, and subsequently dispersed.
An interesting consequence of this scenario is that the stellar component of the dwarf could still be found lurking relatively intact in the halo, since it typically takes multiple orbits for an infalling satellite to lose all of its dark matter and disperse.

SC1 and its associated lower-mass clusters provide new clues to the long-standing mystery of how low-density star clusters are formed.
In this case there appears to be a connection to an external perturbation of the host galaxy, which builds on existing ideas that low-density star clusters may form preferentially in low-density regions during galactic interactions.

Various avenues are available to illuminate the origins of SC1 and of ECs and FFs more generally.
Further spectroscopic analysis and observations
of SC1 and the host galaxy disc could determine if their chemical abundance patterns are the same.
Theoretical modeling is needed of the orbit of SC1, of various interaction scenarios, and of the physics of star cluster formation at low densities.
Observational surveys of the outer regions of star-forming galaxies and tidal debris could determine if and when low-density young star clusters are prevalent.







\section*{Acknowledgements}

We thank the referee for useful comments; Annette Ferguson, Mark Gieles, Oleg Gnedin,  Alison Sills and Jeremy Webb for helpful discussions; 
and Michael Balogh, Zach Jennings, Robert Lupton, Paul Price and Asher Wasserman for technical assistance.
This research is based on observations with the NASA/ESA Hubble
Space Telescope obtained at the Space Telescope Science Institute, which
is operated by the Association of Universities for Research in Astronomy, Incorporated,
under NASA contract NAS5-26555. Support for Program number
HST-GO-14748 was provided through a grant from the STScI under NASA contract
NAS5-26555.
A.J.R.\ was supported by National Science Foundation (NSF) grant AST-1616710 and as a Research Corporation for Science Advancement Cottrell Scholar.
J.L.C.\ acknowledges support from NSF grant AST-1816196.
D.J.S.\ acknowledges support from NSF grants AST-1821967 and 1813708.
S.C.\ was supported by the IBM Einstein Fellowship held at the Institute for Advanced Study, and by NSF AST 2009828.
D.C.\ is supported by NSF grant AST1814208.
C.T.G.\ is supported by NSF grants AST-1615838 and AST-1813628.
This research was supported in part by the National Science Foundation under Grant No.\ NSF PHY-1748958.

Some of the data presented herein were obtained at the W.~M.\ Keck Observatory, which is operated as a scientific partnership among the California Institute of Technology, the University of California and the National Aeronautics and Space Administration. The Observatory was made possible by the generous financial support of the W.~M.\ Keck Foundation. 
The authors wish to recognize and acknowledge the very significant cultural role and reverence that the summit of Maunakea has always had within the indigenous Hawaiian community.  We are most fortunate to have the opportunity to conduct observations from this mountain. 
This research has made use of the NASA/IPAC Extragalactic Database (NED), which is funded by the National Aeronautics and Space Administration and operated by the California Institute of Technology.

\section*{Data Availability}

The {\it HST} and Keck underlying this article are available in MAST and
the Keck Observatory Archive, respectively.
The Subaru and CTIO data underlying this article will be shared on reasonable request to the corresponding author.



\bibliographystyle{mnras}
\bibliography{n247}








\bsp	
\label{lastpage}
\end{document}